\documentstyle{jfm}

\catcode`\œ=\active \gdefœ{\setbox0=\hbox{0}\hbox to\wd0{}}

\ifCUPmtlplainloaded
\else
\fi

\ifCUPmtlplainloaded
  \ifoldfss

  \fi

  \renewcommand{\simeq}{\approx}
\fi

\ifCUPmtlplainloaded
  \font\bit = mtmib10 at 10.5pt \skewchar\bit ='177  % bold math italic
\else
  \font\bit = cmmib10 \skewchar\bit ='177  % bold math italic
\fi

\ifCUPmtlplainloaded
\else
  \font\tenbmi=cmmib10 at 10pt  \skewchar\tenbmi ='177
  \font\sevenbmi=cmmib10 at 7pt \skewchar\sevenbmi ='177
  \font\fivebmi=cmmib10 at 5pt  \skewchar\fivebmi ='177

  \newfam\bmifam
  \textfont\bmifam=\tenbmi
  \scriptfont\bmifam=\sevenbmi
  \scriptscriptfont\bmifam=\fivebmi
  
\fi

\newsavebox{\thalfbox}
\sbox{\thalfbox}{$\textstyle\frac{1}{2}$}

\newsavebox{\shalfbox}
\sbox{\shalfbox}{$\scriptstyle\frac{1}{2}$}

\newsavebox{\squartbox}
\sbox{\squartbox}{$\frac{1}{4}$} %RM removed scriptstyle

\newsavebox{\etbox}
\sbox{\etbox}{\boldmath$\eta$}
\newsavebox{\astrutbox}
\sbox{\astrutbox}{\rule[-5pt]{0pt}{20pt}}

\ifnfsstwo

\fi
\ifnfssone
  \newmathalphabet{\mathit}
    \addtoversion{normal}{\mathit}{cmr}{m}{it}
    \addtoversion{bold}{\mathit}{cmr}{bx}{it}

\fi
\ifoldfss

\fi

\mathchardef\varLambda="0103

\ifCUPmtlplainloaded
\else
\fi

\ifCUPmtlplainloaded
  \let\bcdot=\undefined
  \NewSymbolFont{bldsym}{mtbsy10}{'60}
  \NewMathSymbol{\bcdot}{2}{bldsym}{01}
\else
  \font\tenbms=cmbsy10          \skewchar\tenbms ='60
  \font\sevenbms=cmbsy10 at 7pt \skewchar\sevenbms ='60
  \font\fivebms=cmbsy10 at 5pt  \skewchar\fivebms ='60

  \newfam\bmsfam
  \textfont\bmsfam=\tenbms
  \scriptfont\bmsfam=\sevenbms
  \scriptscriptfont\bmsfam=\fivebms

  \edef\bsy{\hexnumber\bmsfam}
  \mathchardef\bnabla="0\bsy72
  \mathchardef\bcdotsymbol="0\bsy01
  \def\bcdot{\,\bcdotsymbol\,}
\fi

\newtheorem{lemma}{Lemma}

\input epsf

\begin{document}

\title[Intermediate Shocks]{On the Inadmissibility of Non-evolutionary Shocks}

\author[S.A.E.G.  Falle \and  S.S.   Komissarov]{Samuel A.E.G. Falle  \and
Serguei S. Komissarov}

\affiliation{Department  of Applied   Mathematics,  \\  The  University of
Leeds, \\ Leeds LS2 9JT, UK}

\maketitle

\begin{abstract}
In recent  years,  numerical solutions  of the  equations  of compressible
magnetohydrodynamic flows have been  found to contain  intermediate shocks
for certain kinds of  problems.  Since these results  would seem to be  in
conflict  with  the classical  theory of  magnetohydrodynamic shocks, they
have stimulated attempts to  reexamine various aspects  of this theory, in
particular the role of  dissipation.  In this paper  we study  the general
relationship   between    the  evolutionary  conditions  for discontinuous
solutions of the dissipation-free  system and the existence and uniqueness
of steady    dissipative   shock structures   for   systems of quasilinear
conservation laws with a concave entropy function. Our results confirm the
classical theory.  We also show that the appearance of intermediate shocks
in numerical  simulations can be understood in  terms of the properties of
the equations of planar  magnetohydrodynamics   for which some  of   these
shocks  turn out to  be evolutionary.   Finally, we discuss  ways in which
numerical schemes can be  modified  in order to  avoid the  appearance  of
intermediate shocks in simulations with such symmetry.
\end{abstract}

%================================================================
\section{Introduction}
\label{intro}
%================================================================

It is  well known  that  not  all discontinuous solutions  of   hyperbolic
conservation laws  are  admissible.   Some  of these   can  be excluded on
physical grounds.   For example, expansion shocks in  gas dynamics must be
discarded since  they  do not satisfy the  second  law of  thermodynamics.
Others can  be excluded for purely mathematical  reasons  such as the fact
that  they do  not  satisfy  uniqueness and  existence  conditions  or are
structurally unstable  with respect to  small perturbations of the initial
data.  These  mathematical  conditions   are  usually called  evolutionary
conditions.   For  example,  intermediate shocks   in magnetohydrodynamics
(MHD) satisfy the second law but are not evolutionary.

This  subject was  extensively studied between  the  late 1940's and early
1960's (e.g.  Courant \& Friedrichs 1948, Lax 1957, Akhiezer et al.  1959,
Germain 1960, Gel'fand 1963, Polovin 1961) and a full account can be found
in numerous textbooks (e.g.  Jeffrey \& Taniuti 1964, Cabannes 1970, Somov
1994).  Until recently there was general  agreement that admissible shocks
must  both satisfy   the   evolutionary condition  and  possess a   steady
dissipative   shock  structure,  although   the   relation   between these
conditions   was not   entirely clear.   There   the  matter rested  until
time-dependent numerical solutions of the dissipative MHD equations showed
that certain  types of intermediate shocks can  arise  from smooth initial
data (Wu 1987).  Shortly thereafter, Brio  \& Wu (1988) found intermediate
shocks in their numerical  solution for a  particular MHD Riemann problem.
More  recently,    intermediate shocks   have  been  also   been  found in
two-dimensional simulations (De Sterck  et al.  1998).   Furthermore, Chao
et al.  (1993) have reported a detection of an interplanetary intermediate
shock in the Voyager 1 data.   All this has caused  some authors to reject
the classical theory and to suggest that the evolutionary condition is not
relevant to dissipative MHD (Wu  1987, 1988a,b, 1990; Kennel, Blandford \&
Wu 1990; Hada 1994, Myong  \& Roe 1997a,b)  and has led to a reexamination
of the whole  question of  the existence,  or otherwise, of  non-classical
shocks (see Glimm  1988, Freistuhler \& Liu  1993, Myong \& Roe 1997a  and
references  therein).  There are, however,  others who argue that there is
nothing  wrong  with the  classical   theory (e.g.  Barmin,  Kulikovsky \&
Pogorelov 1996; Falle \& Komissarov 1997).

The matter clearly needs to be resolved, particularly since the existence,
or otherwise of intermediate shocks is  of crucial importance not only for
fundamental  MHD processes such  as  reconnection (Wu  1995), but is  also
relevant to many  other  astrophysical applications.  The purpose  of this
paper is to try and clear the matter up by showing that there is neither a
real conflict  between  the classical  shock theory   and the results   of
numerical    calculations  nor   any  incompatibility  between   ideal and
dissipative  MHD. In order  to make  the discussion complete,  we have put
together and  extended a number of results  from the  literature that have
tended be ignored or misunderstood.

This paper is organised as follows.  In \S\ref{sgeneral} we briefly review
the  classical  shock    theory and  the    evolutionary conditions.    In
\S\ref{dis-str} we study the relationship between these conditions and the
uniqueness  and  existence   of steady dissipative  shock   structures for
systems with a concave entropy function.  In  \S\ref{apple} we apply these
results to the full system of MHD  equations and to  the reduced system of
planar  MHD.   In \S\ref{numerics}  we present   the  results of numerical
calculations which show that, for both these systems, the behaviour of the
shocks is entirely consistent with the  predictions of the classical shock
theory.  In  \S\ref{dissc} we consider  various aspects  of the problem of
intermediate shocks and discuss ways in which to avoid their appearance in
MHD simulations  with planar  symmetry.    In particular,  we present  the
results   of  one dimensional simulations   using  a modified Glimm scheme
(Glimm 1965) in which these shocks do not appear.

%================================================================
\section{General Theory of Shocks} 
\label{sgeneral}
%================================================================

In this    section we give  a  brief  review of    the classical theory of
discontinuous solutions of hyperbolic conservation laws.  For our purposes
it is sufficient to consider only the dimensional equations of the form
\begin{equation}
{{\partial {\bf u}} \over {\partial t}} +  {{\partial {\bf f}} \over
{\partial x}} = 0, 
\label{a1}
\end{equation}
where ${\bf u}  \in  {\cal R}^n$ is  a  vector of  conserved variables and
${\bf f}({\bf u}) \in {\cal R}^n$ is a vector of the corresponding fluxes.
 
As  is well known,   the system  (\ref{a1})  is called  hyperbolic if  the
Jacobian matrix
\[
{\bf A} = {{\partial {\bf f}} \over {\partial {\bf u}}}.
\]
has $n$ real eigenvalues, $\lambda_k$ ($k = 1  \ldots n$) corresponding to
$n$   linearly independent right eigenvectors,  ${\bf  r}_k$ and is called
strictly  hyperbolic if all  the  $\lambda_k$ are different.  The physical
significance of  the  $\lambda_k$ is that  they  are the  speeds  of small
amplitude waves.

Waves are classified as linear or nonlinear according to the behaviour of
\[  
C_k({\bf u}) \equiv {\bf r}_k({\bf u}) \cdot \nabla_u \lambda_k({\bf u}).
\]
If $C_k({\bf u})=0$  for all ${\bf u}$, then  the k-wave is called linear,
whereas if the  dimension of  the  surface defined by  $C_k({\bf u})=0$ is
less then $n$, then it is called nonlinear or genuinely nonlinear.

The  states, ${\bf u}_l$,  ${\bf u}_r$  on  either side of a discontinuity
travelling with speed $s$ must satisfy the shock equations
\begin{equation}  
s({\bf u}_l - {\bf u}_r) = {\bf f}_l - {\bf f}_r ,
\label{a4}
\end{equation}  
The number $n_s$ of independent shock equations can  be less then $n$. For
example,  a contact  discontinuity in  gas   dynamics has  $n_s=3$ whereas
$n=5$.   Since ${\bf A}$ is  the Jacobian, we  clearly have $s \rightarrow
\lambda_k$ for some $k$ as ${\bf u}_l \rightarrow  {\bf u}_r$, which means
that one can associate each discontinuity that allows  this limit with one
of the waves  of  the system. A  discontinuity  is  called linear if   the
corresponding characteristic speed does not change across it, otherwise it
is called  nonlinear.   There  mere fact  that a   discontinuity satisfies
(\ref{a4}) it does not necessarily imply that it  is either stable or that
it can arise from continuous initial data.

For some hyperbolic  systems  equations (\ref{a4}) allow  nonlinear shocks
that propagate  with a  characteristic  speed associated with  a nonlinear
wave,  which means that  they  can be  attached  to  such a  wave to  form
compound waves.  Systems with  such shock solutions are called non-convex.
Compound waves may arise from continuous initial data if the system allows
single simple waves  in which $C_k({\bf  u})$ changes sign along the phase
curve of a  simple  wave.  This condition   is therefore often used as  an
alternative definition of non-convexity.   Although these  definitions are
equivalent for a single conservation law, they are  not necessarily so for
systems.
  
The  evolutionary condition  is   directly related   to the   question  of
existence  and  uniqueness of  discontinuous solutions.  It  is well known
that for hyperbolic  equations there  is a general   way of deciding  this
question,   which is to   use the  compatibility conditions  that  must be
satisfied   along   the    characteristics   (Friedrichs  1955).     If  a
characteristic  with  wave  speed   $\lambda_k$   enters one  side   of  a
discontinuity, then the state on that side  must satisfy the compatibility
relation associated with that characteristic,
\[  
  {\bf l}_{k}({\bf u}) \cdot {\bf du}=0,  
\]
where  ${\bf l}_{k}(\bf   u)$  is the   left  eigenvector  of  ${\bf   A}$
corresponding  to  that  characteristic.  These equations  are independent
provided the ${\bf l}_k$ are linearly independent i.e.  for all hyperbolic
systems.  If the wave speeds on either side of  the discontinuity are such
that  $m_i$ compatibility relations have to  be  satisfied, then there are
$n_s+m_{i}$ equations relating the  $2n+1$  unknowns associated with   the
discontinuity, ${\bf   u}_l$, ${\bf u}_r$ and  the   shock speed,  $s$.  A
discontinuous solution can therefore only exist and be unique if
\begin{equation}  
 m_{i} = 2n-n_s+1. 
\label{a5}
\end{equation}  
Obviously,  when  $n_s=n$,  (\ref{a5}) reduces to
\begin{equation}  
 m_{i} = n+1. 
\label{a5a}
\end{equation}  
It is clear  from this that if a  characteristic is parallel to the  shock
curve,  then  it is     counted   as incoming  since  the    corresponding
compatibility relation must be satisfied (Gelfand 1963).

If  $m_{i}>2n-n_s+1$ then  the system is  overdetermined and   there is no
solution except for   certain   special initial conditions.   There   will
therefore always be arbitrarily small perturbations of this data that will
destroy such a discontinuity by splitting it into  a number of waves, just
as an  arbitrary  initial dicontinuity splits  in a  Riemann problem.   If
$m_{i}<2n-n_s+1$,  then  the solution  exists, but is  not unique  and one
might hope that this nonuniqueness can be removed by including dissipative
terms.  In  the  following we will  call condition  (\ref{a5})  the strong
evolutionary condition and call the condition
\[
 m_{i} \le 2n-n_s+1 ,
\]
which allows nonunique solutions a relaxed evolutionary condition.

An equivalent way   of obtaining (\ref{a5})   is  by a linear   structural
stability  analysis of shock solutions   (e.g.  Landau  \& Lifshitz  1959,
Jeffrey  \& Taniuti 1964).   A discontinuity  that is  exposed to  a small
amplitude  incident wave will only survive  if  it can respond by changing
its speed and emitting small amplitude waves.  Each such wave is described
by one parameter and  we also have  the perturbation  in the shock  speed,
which means  that there are  $m_{o}+1$  unknowns  in this  problem,  where
$m_{o}$ is  the  number   of outgoing characteristics.   Since  these  are
related to  the   amplitude  of the incoming  wave   by  the $n_s$   shock
relations, the discontinuity can only have a unique response if
\begin{equation}  
 m_{o} = n_s-1. 
\label{a6}
\end{equation}  
It is  worth pointing out that, contrary  to what is  claimed in  Myong \&
Roe, 1997a, this analysis does not assume that  the discontinuity is weak.
This suggests that non-unique discontinuous solutions should spontaneously
self-destruct by  emitting waves even if they  are not perturbed (Anderson
1963).  Although the  conditions  (\ref{a5}) and  (\ref{a6}) appear  to be
different,  the fact  that  $m_{o}+m_{i}=2n$ means that  they are entirely
equivalent  (Gel'fand  1963).  Note that,   if  the system  of   shock and
compatibility equations   splits into    independent subsets,   then   the
discontinuity is only evolutionary if  each of these  subsets has the same
number of equations as variables (Jeffrey \& Tanuiti 1964).

Finally, as far  as the evolutionary conditions are  concerned it does not
matter  whether, or not, the system  (\ref{a1}) is strictly hyperbolic and
convex since  these   properties  are  not   used  in  the derivation   of
(\ref{a5},\ref{a6}).   However,  it  is  only   in  the case   of strictly
hyperbolic systems that these conditions reduce to the Lax conditions (Lax
1957)
\[
\begin{array}{rcl}  
  \lambda_{k-1}({\bf u}_l) < & s & < \lambda_{k}({\bf u}_l) \\ 
  \lambda_{k}({\bf u}_r) < & s & < \lambda_{k+1}({\bf u}_r) 
\end{array} ,  
\]
for  a nonlinear  discontinuity associated  with  the $k$th characteristic
(here we have assumed that $\lambda_1 < \lambda_2 < ... < \lambda_n$).

%================================================================
\section{Evolutionary conditions and dissipative shock structure}
\label{dis-str}
%================================================================

In  order  to assess  recent  claims  that  nonevolutionary shocks  become
admissible if  dissipative  terms are  included, we need   to  look at the
general relationship    between  the  evolutionary   conditions   and  the
uniqueness and existence of  steady dissipative shock structures.  Godunov
(1961)  has shown that it  is much easier to explore  this question if the
equations can be  transformed to a symmetric  form.  Although this is  not
possible   for arbitrary hyperbolic systems   of conservation laws, it can
certainly be done  for gasdynamics, MHD, and  the shallow water  equations
and probably for any system that can arise in nature.

% ------------------------------------
\subsection{Symmetric Form of the Ideal Equations}
\label{sb-symeq}
% ------------------------------------

We start  by  summarizing  some of   the results  described  by Friedrichs
(1954), Friedrichs  \& Lax (1971) and Boillat  (1974, 1982). As before, it
is only necessary to consider the one dimensional case.

Consider  a dissipation-free system of conservation  laws described by the
equations (\ref{a1}). Suppose now that   there exists a quantity,  $h({\bf
u})$, which is  also conserved as  long as the solution  to this system is
continuous.  For example,  $h({\bf u})$ is  the entropy  in gasdynamics or
MHD, whereas it is the  total energy for the  shallow water equations.  If
such a  quantity exists, then there  must  exist a flux  function, $g({\bf
u})$, such that
\begin{equation} 
   \frac{\partial h}{\partial t} +  
   \frac{\partial g}{\partial x} = 0, 
\label{9}   
\end{equation}  

(\ref{a1}) and (\ref{9}) can  only be consistent if
\begin{equation} 
   \frac{\partial h}{\partial u_i}   
   \frac{\partial f_i}{\partial u_j} = 
   \frac{\partial g}{\partial u_j},  
\label{10}   
\end{equation}  
(summation convention assumed), since then

\[
   \frac{\partial h}{\partial t} +  
   \frac{\partial g}{\partial x} =  
   \frac{\partial h}{\partial u_i} \left(
   \frac{\partial u_i}{\partial t} +  
   \frac{\partial f_i}{\partial x} \right) = 0 . 
\]
for any $C^1$ solution satisfying (\ref{a1}).

If we now use  $h$  to define the Legendre transformation
\begin{eqnarray} 
\label{12}
   u^{\prime}_i & = & -\frac{\partial h}{\partial u_i}, \\
\label{13}
   u_i & = & \frac{\partial h^\prime}{\partial u^\prime_i}, \\
\label{14}
   h^\prime  &=& h + {u^\prime}_i {u_i},
\end{eqnarray}  
then (\ref{10}) allows us to write the fluxes as
\[
f_i = {{\partial g^\prime} \over {\partial u^\prime_i}}
\]
where
\[
  g^\prime = g + {u^\prime}_i f_i.
\]

In terms of the variables ${\bf  u^\prime}$, (\ref{a1}) becomes a symmetric
system

\begin{equation}
   {\bf P} \frac{\partial {\bf u}^\prime}{\partial t} +  
   {\bf Q}\frac{\partial {\bf u}^\prime}{\partial x} = 0
\label{17}   
\end{equation}  
where the symmetric matrices ${\bf  P}$ and ${\bf  Q}$ are given by

\begin{equation}  
\begin{array}{ccccccc}
   P_{ij} & = &
          \displaystyle{\frac{\partial u_i}{\partial u^{\prime}_j}} & = &
          \displaystyle{\frac{\partial^2 h^\prime}
          {\partial {u^\prime}_i \partial u^{\prime}_j}}
          & = &
          \displaystyle{- \frac{\partial^2 h}
          {\partial u_i \partial u_j}}, \\
 & & & & & & \\   
   Q_{ij} & = &
          \displaystyle{\frac{\partial f_i}{\partial {u^\prime}_j}} & = &
                \displaystyle{\frac{\partial^2 g^\prime}
               {\partial {u^\prime}_i \partial {u^\prime}_j}}.\\
\end{array}
\label{18}   
\end{equation}
Note  that  $h$  is usually a   strictly concave  function, in which  case
(\ref{18}) ensures   that  ${\bf   P}$  is  positive  definite   and   the
transformation is non-singular. In ordinary gasdynamics or MHD, $h$ is the
entropy per unit volume and  is therefore guaranteed to  be concave by the
second law of  thermodynamics. For the shallow water  equations $h =  -e$,
where $e$  is the sum of the  kinetic and potential energy and dissipation
ensures that this is also concave.

% ------------------------------------
\subsection{Dissipative Equations}
\label{sb-diss}
% ------------------------------------

If we  now assume  that  the dissipative fluxes  are proportional   to the
spatial gradients of the dependent variables, then the dissipative version
of (\ref{17}) is
\begin{equation} 
   \frac{\partial {\bf u}}{\partial t} +  
   \frac{\partial {\bf f}}{\partial x} = 
   {\bf P} \frac{\partial {\bf u}^\prime}{\partial t} +  
   {\bf Q}\frac{\partial {\bf u}^\prime}{\partial x} =
   \frac{\partial} {\partial x} {\bf D}
   \frac{\partial {\bf u^\prime}}{\partial x}
\label{20}   
\end{equation}  
where ${\bf D}$ is a matrix of dissipation coefficients.  Multiplying this
on the left by ${\bf u}^{\prime t}$ (the superfix t denotes the transpose)
and using (\ref{9}--\ref{12}) gives the evolution equation for $h$
\[
   \frac{\partial h}{\partial t} +  
   \frac{\partial g}{\partial x} = 
   -{\bf u}^{\prime t} \frac{\partial} {\partial x} 
           {\bf D} 
   \frac{\partial {\bf u}^\prime}{\partial x} ,    
\]

Integrating this over an arbitrary  fixed interval $[a,b]$ and integrating
the dissipative term by parts gives
\[
   \frac{d}{d t} \int\limits_a^b h d x +  
   \left[ g
   +{\bf u}^{\prime t} {\bf D} 
   \frac{\partial {\bf u}^\prime}{\partial x} \right]^b_a =   
   \int\limits_a^b    
   \frac{\partial {\bf u}^{\prime t}}{\partial x}     
           {\bf D}  
   \frac{\partial {\bf u}^\prime}{\partial x}  d x .    
\]
Since the  term on the RHS of  this equation represents  a source term for
$h$ and the second law of thermodynamic requires that  this be positive if
$h$ is the entropy per unit volume, the matrix ${\bf  D}$ must be positive
definite for gasdynamics and MHD.  The dissipative shallow water equations
must  also satisfy this condition if   we set $h  = -e$,  where $e$ is the
total energy.

One  can also show  that all linear  waves decay if  ${\bf D}$ is positive
definite  and $h$ is a strictly  concave.  The linear version of (\ref{20})
is simply
\[
   {\bf    P}  \frac{\partial   {\bf u}^\prime}{\partial  t}     + {\bf Q}
   \frac{\partial   {\bf  u}^\prime}{\partial   x}    =  {\bf    D}
   {{\partial^2 {\bf u}^\prime} \over {\partial x^2}}
\]
where ${\bf   P}$, ${\bf Q}$, and  ${\bf  D}$ are now   constant matrices.
Multiplying this  by  ${\bf u}^{\prime t}$  and   integrating over $[a,b]$
gives
\[
{d \over {d t}} \int_a^b {{\bf u}^{\prime t} {\bf P} {\bf u}^\prime d x} +
\left[{\bf  u}^{\prime t} {\bf  Q} {\bf u}^\prime  -  2 {\bf u}^{\prime t}
{\bf D}  \frac{\partial {\bf  u}^\prime}{\partial   x} \right]^b_a =  -  2
\int_a^b {{{\partial {\bf u}^{\prime   t}}  \over {\partial x}}   {\bf  D}
{{\partial {\bf u}^\prime} \over {\partial x}} d x},
\]
after  integrating  the dissipative term  by  parts.  Since ${\bf  P}$  is
positive definite if $h$ is strictly concave,  the term on the RHS ensures
that all linear waves decay if ${\bf D}$ is positive definite,

% ------------------------------------
\subsection{Steady Shock Structures}
\label{sb-shst}
% ------------------------------------

Now consider a solution of  the steady version of (\ref{20})
\begin{equation}
{d \over {d x}} {\bf  f} = {d  \over {d x}}  {\bf D} {d \over {d x}}
{\bf u^\prime}
\label{27}
\end{equation}
with the boundary conditions
\begin{equation}
\begin{array}{l}
{\bf u}^\prime \rightarrow 
\left\{ {
\begin{array}{l}
{\bf u}^\prime_l~~x \rightarrow -\infty \\
{\bf u}^\prime_r~~x \rightarrow +\infty.
\end{array} } \right. 
\end{array} 
\label{28}
\end{equation}
If this  represents a shock  structure, then ${\bf u}^\prime_l$  and ${\bf
u}^\prime_r$ must satisfy the shock relations in the shock frame
\begin{equation}
{\bf f}({\bf u}^\prime_l) = {\bf f}({\bf u}^\prime_r).
\label{29}
\end{equation}

Integrating  (\ref{27})  and applying  the boundary  conditions (\ref{28})
gives
\begin{equation}
{\bf D}  {{d {\bf u}^\prime}  \over {d x} } =  {\bf f}({\bf u}^\prime) -
{\bf   f}({\bf  u}^\prime_l) =   {\bf  f}({\bf  u}^\prime) - {\bf  f}({\bf
u}^\prime_r)
\label{30}
\end{equation}
A steady shock structure therefore corresponds to a solution of (\ref{30})
that  connects  the  equilibrium   points  ${\bf  u}^\prime_l$   and ${\bf
u}^\prime_r$. We now show that there is no guarantee that this solution is
unique  and  structurally  stable  unless  the corresponding discontinuous
solution of the  ideal system are   satisfies the evolutionary  conditions
(\ref{a5}).

Let $L_u$ be  the unstable manifold  of  the point ${\bf u}^\prime_l$  and
$R_s$  the stable  manifold  of the  point ${\bf  u}^\prime_r$.   Then the
trajectories in $L_u$   and $R_s$  are   described by $dim(L_u)  - 1$  and
$dim(R_s) - 1$ parameters respectively. Since any trajectory which lies in
both has  to satisfy  $n - 1$   matching conditions, this means  that,  in
general,  there will   only  be  a  unique   trajectory connecting   ${\bf
u}^\prime_l$ and ${\bf u}_r$ if $dim(L_u) + dim(R_s) = n + 1$. If $dim(L_u)
+ dim(R_s)  > n + 1$,  then the trajectory may not  be unique,  whereas if
$dim(L_u) + dim(R_s) < n + 1$, then any trajectory  that does exist can be
destroyed by  perturbations  of   ${\bf u}^\prime_l$,  ${\bf  u}^\prime_r$
i.e. it is not structurally stable.

The following theorem relates $dim(L_u)$  and $dim(R_s)$ to the number  of
characteristics entering the shock:
\begin{theorem} 
If ${\bf u}^\prime_e$  is an  equilibrium  point of  the dissipative shock
equations (\ref{30}) at  which  none of the characteristic  speeds vanish,
then the equilibrium  point is hyperbolic and the  dimension of its stable
(unstable)  manifold   is  given  by the   number  of positive  (negative)
characteristic speeds in the state ${\bf u}^\prime_e$.
\end{theorem}
\begin{proof}
Suppose that ${\bf u}^\prime_e =  {\bf  u}^\prime_l$ (the proof for  ${\bf
u}^\prime_r$  is   identical).    Then  linearizing   (\ref{30}) in    the
neighbourhood of ${\bf u}^\prime_l$ gives
\[
{\bf D}_l {{d {\bf v}} \over {d x}}  =  {\bf Q}_l{\bf v}, 
\]
where ${\bf  v} = {\bf u}^\prime  -  {\bf u}^\prime_l$, ${\bf  Q}_l = {\bf
Q}({\bf u}^\prime_l)$  and ${\bf  D}_l = {\bf   D}({\bf u}^\prime_l)$.  If
this  equilibrium point is hyperbolic,  then  the dimension  of its stable
(unstable)  manifold  are given  by the numbers  of  eigenvalues, $\mu_k$,
satisfying
\begin{equation} 
 |{\bf Q}_l -\mu {\bf D}_l| =0 . 
\label{32}   
\end{equation}  
with   positive  (negative) real parts.

On the other  hand, the characteristic  speeds for the  system (\ref{17}),
$\lambda_k$, in the state ${\bf u}^\prime_l$ are given by
\begin{equation} 
   \left| {\bf Q}_l - \lambda {\bf P}_l \right| = 0 . 
\label{33}
\end{equation}  
A standard result (e.g. Gantmacher 1959) tells us that, since ${\bf P}_l$,
${\bf Q}_l$  are symmetric and  ${\bf  P}_l$ is  positive definite,  ${\bf
Q}_l$ has the  same number of positive, negative  and zero  eigenvalues as
the set $\lambda_k$.  If, like Godunov  (1961), we assume that ${\bf D}_l$
is symmetric as well  as positive definite, then  the theorem would follow
immediately from (\ref{32}) and  (\ref{33}).  However, the following lemma
shows that this is an unnecessary restriction.
\begin{lemma}
Let ${\bf  Q}$ be a non-singular  symmetric  matrix, ${\bf D}$  a positive
definite matrix and $\mu_k$ the solutions of
\[
 |{\bf Q} -\mu {\bf D}| =0. 
\]
Then the number of $\mu_k$ with positive (negative)  real part is the same
as the number of positive (negative) eigenvalues of ${\bf Q}$.
\label{lem1}
\end{lemma}
\begin{proof}
Define
\[
   {\bf D}_\epsilon = {\bf D}_s + \epsilon {\bf D}_a,
\]
where $\epsilon \in [0,1]$ and   
\[
    {\bf D}_s = \frac{1}{2}({\bf D}+{\bf D}^t) ,\quad 
    {\bf D}_a = \frac{1}{2}({\bf D}-{\bf D}^t) .  
\]
It easy to see that ${\bf D}_\epsilon$ is also positive definite.  

Now consider the eigenvalue problem  
\[
 |{\bf Q} -\mu(\epsilon) {\bf D}_\epsilon| =0 . 
\]
The conclusion of the  lemma is certainly true  for $\epsilon = 0$,  since
then ${\bf D}_\epsilon$    is  symmetric.  If   we   can  show that    the
$\mu_k(\epsilon)$  are   continuous  functions   of  $\epsilon$ and   that
$\Re\{\mu_k(\epsilon)\}  \not = 0   ~\forall k$ for $\epsilon \in  [0,1]$,
then it will also be true for $\epsilon = 1$.

The  $\mu_k(\epsilon)$ are the  roots of a polynomial  of degree $n$ whose
coefficients are  polynomials in  $\epsilon$.   A root can  therefore only
change discontinuously by going  to infinity, which can  only occur if the
coefficient, $|D_\epsilon|$, of the  highest   power of $\mu$    vanishes.
However,  this cannot happen  since  $D_\epsilon$ is positive definite for
$\epsilon \in [0,1]$.   The $\mu_k(\epsilon)$ must therefore be continuous
functions of $\epsilon$ for $\epsilon \in [0,1]$.

In  order to   prove that  the  $\mu_k$  cannot cross the  imaginary axis,
suppose  that  for some $k$, $\mu_k(\epsilon)  =  i \eta$, where $\eta$ is
real. If ${\bf a}+i{\bf b}$ is the corresponding eigenvector, we have
\[
\begin{array}{rcl}
   {\bf Qa}+\eta{\bf D}_\epsilon {\bf b} &=& 0 , \\ 
   {\bf Qb}-\eta{\bf D}_\epsilon {\bf a} &=& 0 .
\end{array}
\] 
Multiplying the first of  these by ${\bf  b}^t$. the second by ${\bf a}^t$
and substracting gives
\[
   \eta ( {\bf b}^t{\bf D}_\epsilon{\bf b} + 
    {\bf a}^t{\bf D}_\epsilon{\bf a} ) = 0 .   
\]
Since ${\bf D}_\epsilon$ is positive  definite this requires $\eta =0$ and
hence $\mu_k = 0$, which cannot be true if  the eigenvalues of $\bf Q$ are
non-zero. This completes the proof of the lemma.
\end{proof}
(\ref{32}), (\ref{33})  and lemma  (\ref{lem1})  show that the theorem  is
true even if ${\bf D}$ is not symmetric.
\end{proof}

This is  a somewhat  more direct  proof of a   result which has  also been
obtained  by Kulikovsky \& Lyubimov  (1965).  In their analysis of viscous
shock structures, Myong \& Roe (1997a)  assumed that Theorem 3.1 holds for
MHD, but did not give a proof.

This  analysis tells  us  that if the  shock  relations  (\ref{29}) have a
solution such  that none of the characteristic  speeds given by (\ref{33})
vanish in both  the left  and  the right state  and $m_i$  is the  number of
characteristics entering the shock, then

\begin{enumerate}  
\item  
for $m_i = n+1$ the shock can  have a unique structurally stable dissipative
structure;
\item   
for $m_i > n+1$ the dissipative structure is not guaranteed to be unique.
\item  
for $m_i < n+1$ there might be a unique  dissipative structure but it cannot
be structurally stable.
\end{enumerate}

These conditions are not only compatible with the evolutionary conditions,
they are  complementary to them.  Shocks  for which $m_i  > n  + 1$ have a
dissipative shock structure and  could therefore be regarded as admissible
on these grounds.  However, the left and right states  of such shocks must
be carefully tuned  since they  cannot adjust  themselves to an  arbitrary
small  perturbation of their left  and right states.   Shocks that satisfy
the  relaxed  evolutionary   condition, $m_i   < n  +   1$, are apparently
permitted by  the  ideal  equations,  but cannot establish   a dissipative
structure and must   spontaneously self-destruct.  It   is therefore clear
that  the only physically admissible  shocks are those  those that satisfy
the strong evolutionary conditions (\ref{a5}) or (\ref{a5a}).

Theorem 3.1 gives  us no information  in those cases  for which the  shock
speed   coincides with at least  one   of the  characteristic speeds.  The
corresponding critical  point is  then no longer  hyperbolic  and its type
depends on the details of the particular system.

%================================================================
\section{Application to Magnetohydrodynamics} 
\label{apple}
%================================================================

As we shall see, the mathematical properties of the full system of MHD and
the reduced planar system of MHD are somewhat different and this has to be
clearly    understood       when  the      evolutionary    conditions  are
applied. We therefore discuss these systems separately.
 
% ------------------------------------
\subsection{Full System of MHD}      
\label{sb-full}
% ------------------------------------

It is well known that the one dimensional equations of  MHD can be written
in the form (\ref{a1})  (e.g. Brio \&  Wu 1988).  The conserved quantities
${\bf u}$ and the corresponding fluxes ${\bf f}$ are
\[
\begin{array}{ll}
{\bf u} =
\left[{
\begin{array}{c}
\rho \\
\rho v_x \\
\rho v_y \\
\rho v_z \\
e \\
B_y \\
B_z
\end{array}
}\right]
&
{\bf f} =
\left[{
\begin{array}{c}
\rho v_x \\
\rho v_x^2 + p_g + B^2/2 - B_x^2 \\
\rho v_x v_y - B_x B_y \\
\rho v_x v_z - B_x B_z \\
\{e + p_g + B^2/2 \}v_x - B_x({\bf v}.{\bf B}) \\
v_x B_y - v_y B_x \\
v_x B_z - v_z B_x
\end{array}
}\right]
\end{array}.
\]
Here $p_g$ is the gas pressure,
\[
e = i + {1 \over 2} B^2 + {1 \over 2} \rho v^2 
\]
is the  total energy  per  unit volume and $i$   is the enthalpy  per unit
volume.  Here we use units such that the  velocity of light and the factor
$4\pi$ do not appear.

As  we have already discussed, ideal  MHD has a supplementary conservation
law representing the conservation   of thermodynamic entropy.   The second
law of thermodynamics guarantees that the function $h = \rho S$, where $S$
is  the entropy per   unit mass,  is  strictly concave  (e.g.  TerHaar  \&
Wergeland 1966) and hence that the matrix  ${\bf P}$ defined by (\ref{18})
is  positive  definite.  The system  of  MHD  equations  can  therefore be
written in the symmetric form (\ref{17}) and is hyperbolic.  Although this
has been demonstrated for  relativistic MHD by  Ruggeri \& Strumia (1981),
we have been unable  to find an account  of the corresponding analysis for
classical   MHD  in the literature.  However,    since the derivations are
similar  to those for   the relativistic case,  we  shall simply  give the
symmetric variables. They are
\[
\begin{array}{rclrclrclrcl}

u_1^\prime & = & 
  \displaystyle{\frac{1}{T}\left({w\over \rho} - {1 \over 2}v^2 \right)}, 
& u_2^\prime & = &  \displaystyle{\frac{v_x}{T}}, 
& u_3^\prime & = &   \displaystyle{\frac{v_y}{T}}, 
& u_4^\prime & = &   \displaystyle{\frac{v_z}{T}}, \\
 & & & & & & & & & & & \\
  u_5^\prime & = & \displaystyle{-\frac{1}{T}},  
& u_6^\prime & = &   \displaystyle{\frac{B_y}{T}},
& u_7^\prime & = & \displaystyle{\frac{B_z}{T}}.
\end{array}
\]

There  is  no need to  verify that  the matrix, ${\bf  D}$, of dissipation
coefficients is positive definite, since this must  be true for any system
that  obeys the second law of  thermodynamics.  Indeed,  this condition is
used to derive the dissipative  equations in the  first place (e.g. Landau
\& Lifshitz 1960).  The exact form of symmetrized equations  is also of no
importance for our purposes.  Their existence, does,  however, allow us to
apply the conclusions of the general theory  described in Sections 2 and 3
to dissipative MHD.

% ------------------------------------
\subsubsection{Characteristic Wave Speeds}
% ------------------------------------

Since there are  seven  variables in this  system,  there are seven  waves
whose speeds  are 

\begin{tabular}{lrcl}
 & & & \\
Fast Waves & $\lambda_{f\mp}$ & $=$ & $v_x \mp c_f$, \\
Alfv\'en Waves & $\lambda_{a\mp}$ & $=$ & $v_x \mp c_a$, \\
Slow Waves &  $\lambda_{s\mp}$ & $=$ & $v_x \mp c_s$, \\
Entropy Wave & $\lambda_e$ & $=$ & $v_x$, \\
 & & &
\end{tabular}

\noindent
where   the alfv\'en speed,  $c_a$, and  the slow  and fast speeds, $c_s$,
$c_f$ are given by
\[
c_a = |B_x|  / \surd \rho,
\]
\[ 
c_{s,f}^2 = {1   \over 2} \left[ {a^2  + {B^2
\over  \rho} \mp  {\left\{ {{\left({a^2 +   {B^2 \over  \rho}}\right)}^2 -
{{4a^2B_x^2} \over \rho}} \right\} }^{1/2}}\right], 
\]
where $a$ is the adiabatic  sound speed.  Note that $0  \leq c_s \leq  c_a
\leq  c_f$.  If  $B_x=0$   then  $c_s=c_a=0$,  whereas if  the  transverse
component  of the magnetic field, ${\bf  B}_t$, vanishes then $c_f=c_a$ if
$c_a>a$,   $c_s=c_a$ if $c_a<a$  and  $c_s=c_f=c_a$  if  $c_a=a$. The  MHD
equations are therefore  not strictly hyperbolic.   Brio \& Wu (1988) also
argued that they are non-convex, but we  shall postpone discussion of this
until later.

% ------------------------------------
\subsubsection{Shock Types}
% ------------------------------------

The MHD  shock equations allow two  linear  solutions and several distinct
types of nonlinear solutions which  satisfy the entropy principle that the
entropy  of    a fluid element always   increases.    A convenient  way of
classifying  these is to   use the  jump  in  the transverse component  of
magnetic field, ${\bf B}_t$.  From  the shock equations one finds (Jeffrey
\& Taniuti 1964)
\begin{equation}
[{\bf B}_t(c_a^2 - v_x^2)]_l = [{\bf B}_t(c_a^2 - v_x^2)]_r,
\label{7}
\end{equation}
where $v_x$ is the velocity in the shock frame. Note that if $c_a^2-v_x^2$
does not vanish, then ${\bf B}_t$ on one side of the discontinuity must be
either parallel or anti-parallel to that on the other.

The nonlinear solutions are 

\begin{enumerate}

\item
Slow/Fast shocks, which have non-zero ${\bf B}_t$ in the same direction on
both sides. (\ref{7})  then implies that   there is no  change  in sign of
$(c_a^2 -  v_x^2)$.  The  magnitude of  magnetic  field is  larger  on the
downstream side for fast shocks and smaller downstream for slow shocks.

\item
Intermediate shocks, which also have non-zero  ${\bf B}_t$ but in opposite
directions  on either side  of the  shock  (Anderson 1963, Cabannes 1970).
(\ref{7}) then implies that $(c_a^2 - v_x^2)$ changes sign.

\item
Switch-on shocks, which  have vanishing ${\bf  B}_t$ upstream.   (\ref{7})
then implies that $v_x^2=c_a^2$ on the downstream side.

\item
Switch-off shocks, which have vanishing ${\bf B}_t$ downstream.  (\ref{7})
then implies that $v_x^2=c_a^2$ on the upstream side.

\end{enumerate} 

The linear discontinuities  are 

\begin{enumerate} 
\item 
Alfv\'en discontinuities, which  have $v_x^2=c_a^2$  on both sides.   case
(\ref{7}) then allows an arbitrary change in the direction of ${\bf B}_t$.
However,  the magnitude  of ${\bf B}_t$  remains unchanged,   which is why
these are sometimes called rotational discontinuities.

\item 
Contact discontinuities, which have the same value of $v_x$ on both sides,
but  $v_x^2 \not= c_a^2$.  (\ref{7})   then requires  that ${\bf B}_t$  be
continuous unless  $B_x =  0$ and  the other  shock conditions require all
other variables, except for the density, to be continuous.
\end{enumerate}  

We shall  also  find occasion  to use   the following  classification   of
nonlinear MHD shocks, which  is due to Germain (1960).   The states in the
shock frame are divided into four types
\[
\begin{array}{ll}
{\rm 1)}~~|v_x| > c_f, & {\rm 2)}~~c_f > |v_x| > c_a, \\
{\rm 3)}~~ c_a > |v_x| > c_s, & {\rm 4)}~ c_s > |v_x|,
\end{array}
\]
and a shock is defined to be of type $m \rightarrow n$ if the upstream and
downstream states  are of  types $m$  and $n$ respectively.   From the MHD
shock equations, one  finds   that pressure and specific  volume,   $\tau$
($\tau = 1/\rho$), on each side of a nonlinear shock satisfy the following
equations
\[
   p + G^2 \tau + {1 \over 2} \frac{F_y^2}{(\tau-\tau_a)^2} = F_x ,
\]
\[
   w\tau + {1 \over 2} G^2 \tau^2 + {\tau \over 2 \tau_a }  
        \frac{F_y^2}{(\tau-\tau_a)^2} = H ,
\]
where $G$ is the  mass flux, $F_x$,  $F_y$, $H$  are shock invariants  and
$\tau_a= B_x^2/G^2$.  The analysis in Anderson  (1963) can be used to show
that the function  $H(\tau)$ is as shown in  Figure 1.   $\tau-\tau_i$ has
the same  sign as $v_x^2-c_i^2$, where  $i=s,a,f$.  One can see that there
are six different types of compressive shocks: fast shocks ($1 \rightarrow
2$), slow  shocks ($3 \rightarrow 4$),   and four intermediate  shocks: $1
\rightarrow 3$,  $1 \rightarrow 4$, $2  \rightarrow 3$, and $2 \rightarrow
4$.  Depending  on the relative position  of the maxima  of $H$, there are
also  limit shocks which  propagate  with the fast   speed relative to the
upstream state and/or the slow speed relative to the downstream state (see
figures 1{\it b,c}).   We shall denote  such such shocks by $f \rightarrow
n$ and $n \rightarrow s$  respectively.  These shocks turn  out not to  be
evolutionary, but if they were, then MHD would be a non-convex system.

%fffffffffffffffffffffffffffffffffffffffffffffffffffffffffffffff
\begin{figure} 
\begin{center} 
\leavevmode 
%\hbox{% 
\epsffile[24 26 378 137]{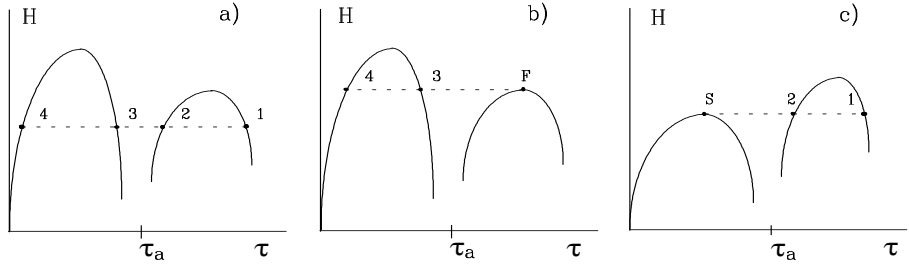}
\caption{The  shock invariant, $H$,   as a  function of  specific  volume,
$\tau$, for three different cases.}
\end{center}
\label{curve-figure}
\end{figure}
%fffffffffffffffffffffffffffffffffffffffffffffffffffffffffffffff 

% ------------------------------------
\subsubsection{Evolutionary conditions} 
% ------------------------------------

When we apply  the evolutionary conditions  to MHD discontinuities we have
to take into account the  fact that the  system of shock and compatibility
equations  split  into    two independent   subsets  for  all     types of
discontinuities,  except the alfv\'en    discontinuity,.  If we choose   a
reference frame  such  that on one   side of  a discontinuity  $B_z=0$ and
$v_z=0$  then    the system  of shock   equations   contains two equations
involving $B_z$ and $v_z$.  These are
\[
  {B_z}_l = {B_z}_r ,
\]
\[
  {v_z}_l = {v_z}_r . 
\]
The  compatibility relations    along the   alfv\'en characteristics  only
involve  $B_z$ and $v_z$ and   they also the only   ones  that do so.   An
evolutionary discontinuity that  is  not  an alfv\'en  discontinuity  must
therefore not only satisfy the general condition (\ref{a5}), but also have
exactly two incoming,   and hence two  outgoing alfv\'en  characteristics.
These  conditions  also    follow from  the   linear  stability   analysis
(Syrovatskii 1959, Jeffrey \& Tanuiti 1964).

In the rest of this section we simply state the well  known results on the
evolutionary   properties of MHD  discontinuities.   We  do, however,  pay
particular  attention  to those cases in   which there are characteristics
travelling with  the same speed as the  discontinuity. As we  have pointed
out in Section 2, such characteristics must be counted as incoming.

There  is no  dispute  about  the  fact that  fast  and  slow shocks   are
evolutionary because they  have  eight  incoming characteristics, two   of
which are   alfv\'en waves.  Furthermore, since  their  speed can never be
equal to a characteristic speed, Theorem 3.1  tells us that they also have
a unique structurally stable dissipative structure.

All intermediate shocks are super-alfv\'enic with  respect to the upstream
state and sub-alfv\'enic with respect to the downstream state, which means
that they  have too many ($>2$)  incoming alfv\'en  characteristics.  They
are    therefore  nonevolutionary and  can   be  destroyed by interactions
alfv\'en waves.

The same argument applies  to switch-on and  switch-off shocks which  also
have  too many  (9) incoming  characteristics,  3  of which are   alfv\'en
characteristics.  However, these solutions  are clearly limits of fast and
slow shocks and  therefore have evolutionary  solutions in their immediate
neighbourhood, which is  why  Jeffrey \&  Taniuti (1964)  call them weakly
evolutionary.  That they   are   not strictly  evolutionary  can  also  be
understood  from the    following example.  Consider   a  switch-on  shock
overtaking  weak  switch-off fast   rarefaction   travelling in the   same
direction.  Once these  have merged,  the  shock is no longer  propagating
into a  state with zero  transverse  magnetic field.   Since the shock  is
superfast, it  has no way  of modifying its  upstream  state and therefore
cannot remain a switch-on shock. Instead, such an interaction leads to the
appearance of a neighbouring fast shock solution, together with some other
waves, at least one of which must, in general, be an alfv\'en wave.

If we count the  two entropy characteristics  as  incoming on the  grounds
that   they   have the  same  speed   as  the  discontinuity, then contact
discontinuities have  eight  incoming characteristics,  two  of  which are
alfv\'en characteristics. They are therefore evolutionary.

Alfv\'en discontinuities  also have  eight incoming characteristics  if we
include the  two alfv\'en characteristics that  have the same speed as the
discontinuity.  The total  number of incoming  alfv\'en characteristics is
three, but this  is allowed since  the fact that  the  shock equations for
these discontinuities couple the $y$  and  $z$ components of velocity  and
magnetic  field  means that this  is  the  one case   for  which the shock
equations do not decompose into two sets.

Theorem   3.1 cannot be  applied  to  contact and alfv\'en discontinuities
since they propagate with a characteristic  speed.  However, they would in
any case  not possess a steady  dissipative structure  simply because they
are   linear and therefore  have no   nonlinear  steepening to balance the
spreading due  to dissipation. For  this reason, Wu (1988b) considers them
to be  inadmissible, but since their  width grows like  $t^{1/2}$, whereas
the separation between the waves in a Riemann problem grows like $t$, they
must be regarded as admissible components of the solution for large times.

% ------------------------------------
\subsection{Reduced system of planar MHD} 
\label{sb-planar}
% ------------------------------------

In this section we discuss the system of  equations which describes MHD in
a world in which the plane defined by  the velocity and the magnetic field
is invariant.  There are several reasons  for doing this.  Firstly, it has
some interesting  properties.  Secondly, we want to  show that the general
classical theory  of shocks is as valid  for this system  as it is for the
full system.  Finally, the  numerical simulations  that  gave rise  to the
current conroversy surrounding intermediate  shocks reflect the properties
of this system.
  
When  the z  components  of the magnetic   field and  velocity vanish, the
equations reduce to  a system of  $5$ variables with the following vectors
of conserved quantities and fluxes
\[
\begin{array}{ll}
{\bf u} =
\left[{
\begin{array}{c}
\rho \\
\rho v_x \\
\rho v_y \\
e \\
B_y \\
\end{array}
}\right]
&
{\bf f} =
\left[{
\begin{array}{c}
\rho v_x \\
\rho v_x^2 + p_g + B^2/2 - B_x^2 \\
\rho v_x v_y - B_x B_y \\
\{e + p_g + B^2/2 \}v_x - B_x({\bf v}.{\bf B}) \\
v_x B_y - v_y B_x \\
\end{array}
}\right]
\end{array}.
\]
This is still a hyperbolic  system but it  is fundamentally different from
the full system of MHD because it does not  have alfv\'en waves.  However,
the   other   characteristic fields  are still   present   with   the same
eigenvalues and with eigenvectors that are the same apart from the reduced
number of components.   Moreover, it has the  same solutions of  the shock
equations including the alfv\'en discontinuity, except  that these are now
only allowed to  change the direction  of the transverse magnetic field by
$\pi$. This follows from the remarkable property of the full system of MHD
that  there  exists an  inertial  frame in    which the variation  of  the
transverse components  of the magnetic field  and  velocity induced by all
characteristic waves  and shocks, except  for alfv\'en waves, are confined
to single  plane.  Note that  the alfv\'en  discontinuity still propagates
with the alfve\'n  speed, but this is  no longer on  of the characteristic
speeds.    The   Riemann problem  for this  system    has been analysed in
considerable detail  by Myong \& Roe  (1997b)  who came to  the conclusion
that the classical evolutionary conditions are inadequate for this system.
However,   we intend to show that   this claim is  based   on a failure to
recognise  the essential difference  between the  reduced  system and full
MHD.
  
% ------------------------------------
\subsubsection{Evolutionary conditions} 
% ------------------------------------

Since the  number of equation  is  reduced by two  and it  is the alfv\'en
waves that are lost, we can conclude that all evolutionary discontinuities
that have two incoming alfv\'en characteristics in  the full system remain
evolutionary in the planar   system.   This implies  that fast,  slow  and
contact discontinuities are evolutionary.

On  the other hand,  discontinuities   that are  evolutionary in  the full
system,   but  which  do    not   have  exactly  two   incoming   alfv\'en
characteristics  must be non-evolutionary in   the planar system. There is
only one  such discontinuity, the  alfv\'en discontinuity, which  now only
has  $5$  incoming characteristics   and  should  therefore  spontaneously
self-destruct even if it is not  perturbed.

Another  interesting  feature   is  that some  of  the    shocks that  are
non-evolutionary  in the  full system  become evolutionary  in the reduced
system.     $1\rightarrow 3$ shocks   now  satisfy the strong evolutionary
condition,    in  fact   they   have  the    same  incoming  and  outgoing
characteristics    as  fast and  switch-on    shocks.     As far  as   the
characteristic  count is   concerned  these  three shocks are    therefore
indistinguishable  so that one  can  use a   single name, {\it  plane fast
shock}, say, for    all of them.   Similarly,  $2  \rightarrow 4$  shocks,
switch-off shocks  and slow shocks become   slightly different versions of
evolutionary {\it plane slow shocks}.

However,   $1 \rightarrow 4$  shocks remain   non-evolutionary even in the
plane system since  they have $7$  incoming characteristics.  Such shocks,
which   have  too many incoming  characteristics,   are  often called {\it
overcompressive} in  the literature.  As we  have  shown, although they do
have a steady dissipative structure, it is not unique and it does not help
them to survive interactions with external perturbations.

$2 \rightarrow  3$ shocks have  only $5$  incoming characteristics and are
therefore non-evolutionary.  Such shocks,    which have too few   incoming
characteristics,  are often called  {\it undercompressive}.  Since they do
not have a  structurally stable steady  dissipative  structure they should
disintegrate spontaneously even without any external perturbation.

Now consider shocks that propagate at one of  the characteristic speeds in
either the   upstream   or downstream   state.   $1  \rightarrow    s$, $f
\rightarrow  4$, and $f \rightarrow  s$  shocks are non-evolutionary since
they have $7$ incoming characteristics.  On the other hand, $2 \rightarrow
s$ and $f \rightarrow 3$ shocks  have $6$ incoming characteristics and are
therefore  evolutionary.  The planar  system of MHD is therefore genuinely
non-convex and admits the  two  evolutionary compound  waves: a  {\it slow
compound wave} consisting  of a $2 \rightarrow  s$ shock  with an attached
slow rarefaction and  a {\it  fast compound   wave} consisting of  a  fast
rarefaction with an attached $f \rightarrow 3$ shock.

Finally, we list the evolutionary shocks  and compound waves of the planar
system along with the notation used in Myong \& Roe (1997b):
\begin{description} 
\item{Slow planar shock} (S1); 
\item{Fast planar shock} (S2); 
\item{Slow compound wave} (C1); 
\item{Fast compound wave} (C2); 
\item{Contact discontinuity} (not considered).  
\end{description} 
Myong \& Roe (1997b) found that some Riemann problems only have a solution
if non-evolutionary   shocks are  permitted.   However,  as  we discuss in
\S\ref{dissc}, these Riemann problems are confined to regions of parameter
space  with zero volume,  which is exactly what is  meant by the statement
that non-evolutionary shocks are structurally unstable.

In the next section we show that the results of numerical calculations are
entirely consistent with these conclusions.

%==========================================================
\section{Numerical Calculations}
\label{numerics}
%========================================================== 

The numerical calculations were carried out  using the scheme described in
Falle, Komissarov  \& Joarder (1998).   This is an  upwind shock capturing
scheme which is capable of dealing with  shocks of arbitrary strength even
without the inclusion of any dissipation other than that introduced by the
truncation errors. Careful  test simulations have  shown that  this scheme
provides accurate solutions  for all  types of  MHD waves  in all regimes.
One can argue  that if a numerical   scheme works well  then its numerical
dissipation must have the   same  qualitative properties as the   physical
dissipation.  However,  in  order to remove any  doubts,  we  modified our
scheme so that it can now handle  dissipative MHD and all the calculations
described here  have a fully  resolved dissipative shock structures (about
15 mesh points wide).   For  this we  used a simple   scalar form for  the
dissipation for which equations (\ref{a1}) become
\[
{{\partial {\bf u}} \over {\partial t}} +  {{\partial {\bf f}} \over
{\partial x}} = {\partial \over {\partial x}} {\bf g},
\]
where the diffusive fluxes are 
\[
{\bf g} = 
\left({
\begin{array}{c}
0 \\
 \\
\displaystyle{{{4 \mu} \over 3}{{\partial v_x} \over {\partial x}}} \\
 \\
\displaystyle{\mu {{\partial v_y} \over {\partial x}}} \\
 \\
\displaystyle{\mu {{\partial v_z} \over {\partial x}}} \\
 \\
\displaystyle{{{4 \mu v_x} \over 3} {{\partial v_x} \over {\partial x}}
+ \mu v_y{{\partial v_y} \over {\partial x}}
+ \mu v_z{{\partial v_z} \over {\partial x}}
+ \nu_m \left[ {B_y {{\partial B_y} \over {\partial x}}
+ B_z {{\partial B_z} \over {\partial x}}} \right]} \\
 \\
\displaystyle{\nu_m {{\partial B_y} \over {\partial x}}}\\
 \\
\displaystyle{\nu_m {{\partial B_z} \over {\partial x}}}\\
\end{array}
} \right)  
\]
where $\mu$ is the dynamic viscosity, $\kappa$ the thermal conductivity
and $\nu_m$ the resistivity.

As expected, the outcomes  of all the simulations  presented here did  not
not depend  on the size of dissipation  and were the  same even  when only
numerical and/or artificial dissipation was  present.  The only effect  of
changing the  dissipation was  to  alter the form  and  width of the shock
structures.

\begin{table}
\begin{center}
\begin{tabular}{l}
$2 \rightarrow 3$ Intermediate Shock (figure 2a)\\
~~Left  state:  $\rho  = 1,~p_g  =  1,~{\bf  v} =  (-0.95,0,0),~{\bf  B} =
 (1,0.5,0)$ \\
~~Right state: $\rho = 0.837,~p_g = 0.705,~{\bf v} = (-1.135,1.266,0),~{\bf
 B} = (1,-0.7,0)$ \\
 \\
Alfv\'en Shock (figure 2b)\\
~~Left  state:  $\rho  = 1,~p_g  =  1,~{\bf  v} =  (-1,1,0),~{\bf  B} =
 (1,1,0)$ \\
~~Right state: $\rho = 1,~p_g = 1,~{\bf v} = (-1,3,0),~{\bf
 B} = (1,-1,0)$ \\
 \\
$1 \rightarrow 3$ Intermediate Shock (figures 3a, 5 and 7a) \\
~~Left  state:  $\rho =   1,~p_g =  1,~{\bf  v} =  (-0.925,0,0),~{\bf B} =
(1,0.5,0)$ \\
~~Right  state: $\rho  = 0.498,~p_g  = 0.258,~{\bf  v} = (-1.857,0.648,0),
~{\bf B} = (1,-0.1,0)$ \\
 \\
$2 \rightarrow 4$ Intermediate Shock (figures 3b and 7b )\\ 
~~Left  state: $\rho   = 1,~p_g =    1,~{\bf v}  =  (-0.4,0,0),~{\bf  B} =
 (0.5,0.5,0)$ \\
~~Right state: $\rho = 0.561,~p_g = 0.155,~{\bf v} = (-0.714,2.252,0),~{\bf
 B} = (0.5,-1.3,0)$ \\
 \\
$1 \rightarrow 4$ Intermediate Shock (figure 4)\\
~~Left state: $\rho = 1,~p_g  = 1.2,~{\bf v}  = (-0.842,0.0,0.0),~{\bf B}  =
 (1.0,0.4,0)$ \\
~~Right state: $\rho = 0.390,~p_g = 0.161,~{\bf v} = (-2.16,0.644,0),~{\bf
 B} = (1.0,-0.142,0)$ \\
 \\
Brio \& Wu Problem (figure 8)\\
~~Left state: $\rho  = 1,~p_g = 1,~{\bf v}  = (0,0,0),~{\bf  B} = (0.75,1,
0)$ \\
~~Right state:  $\rho = 0.125,~p_g =  0.1,~{\bf  v} =  (0,0,0), ~{\bf B} =
(0.75,-1,0)$ \\
 \\
\end{tabular}
\end{center}
\caption{Riemann problems for the numerical calculations.} 
\end{table}

\begin{table}
\begin{center}
\begin{tabular}{llllll} 
Problem & Domain & $n$ & $\mu /\rho $ & $\kappa/\rho$ & $\nu_m$ \\ 
\hline 
\hline 
Figure 2a & $[-4,1]$ & 250 & 0.02 & 0.01 & 0.01 \\
Figure 2b & $[-2,1]$ & 150 & 0.02 & 0.01 & 0.01 \\
Figure 3a,b & $[-4,1]$ & 250 & 0.02 & 0.01 & 0.01 \\
Figure 4 & $[-4,1]$ & 250 & 0.02 & 0.01 & 0.01 \\
Figure 5 & $[-1,1]$ & 200 & 0.01 & 0.005 & 0.005 \\
Figure 6 & $[-2,1]$ & 300 & 0.01 & 0.005 & 0.005 \\
Figure 7a & $[-8,2]$ & 500 & 0.02 & 0.01 & 0.01 \\
Figure 7a & $[-14,1]$ & 750 & 0.02 & 0.01 & 0.01 \\
Figure 8a,b & $[2.5,4.5]$ & 200 & 0.0 & 0.0 & 0.0 \\
\hline 
\end{tabular}
\end{center}
\caption{Other   parameters  for the  numerical  calculations.  $n$ is the
number of mesh points, $\mu$  is the kinematic  viscosity, $\kappa$ is the
thermal conductivity, $\nu_m$ is the resistivity.}
\end{table}

First of all, we need to establish  whether the behaviour of numerical MHD
shocks agrees  with the predictions of  the evolutionary theory.  In order
to do  this, we adopt the  following  procedure.  First  we test whether a
shock has   a steady  dissipative structure  by  setting up   the relevant
Riemann problem and  running the calculation  until a well resolved steady
dissipative shock structure  is established, as expected  for evolutionary
and overdetermined shocks, or a  completely different solution emerges, as
expected  for underdetermined shocks.  If  a steady structure exists, then
we test to see  whether it can  survive small perturbations.  This can  be
accomplished by considering  a slightly different  Riemann problem, as  in
Barmin et al.  (1996) or, like Wu (1988a), allowing a small amplitude wave
to interact with the shock.

%fffffffffffffffffffffffffffffffffffffffffffffffffffffffffffffff
\begin{figure} 
\begin{center} 
\leavevmode 
\hbox{% 
\epsffile[15 13 380 290]{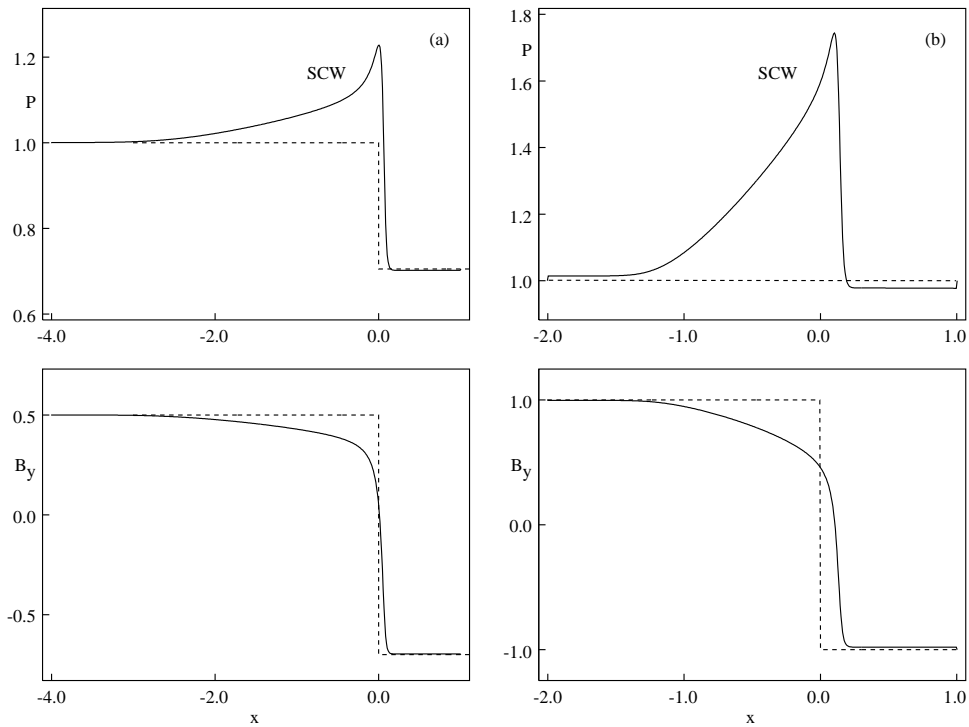}}
\caption{Planar simulations  of shocks  that  should  not have   a  steady
dissipative structure  in  planar MHD: (a) $2   \rightarrow 3$  shock, (b)
Alfv\'en  shock.  In both cases   the  outcome is   a  slow compound  wave
(SCW). The  dashed  lines show the corresponding   initial solutions.  The
continuous lines show the final solutions.}
\end{center}
\label{noshst}
\end{figure}
%fffffffffffffffffffffffffffffffffffffffffffffffffffffffffffffff  

%fffffffffffffffffffffffffffffffffffffffffffffffffffffffffffffff
\begin{figure} 
\begin{center} 
\leavevmode 
\hbox{% 
\epsffile[15 13 380 432]{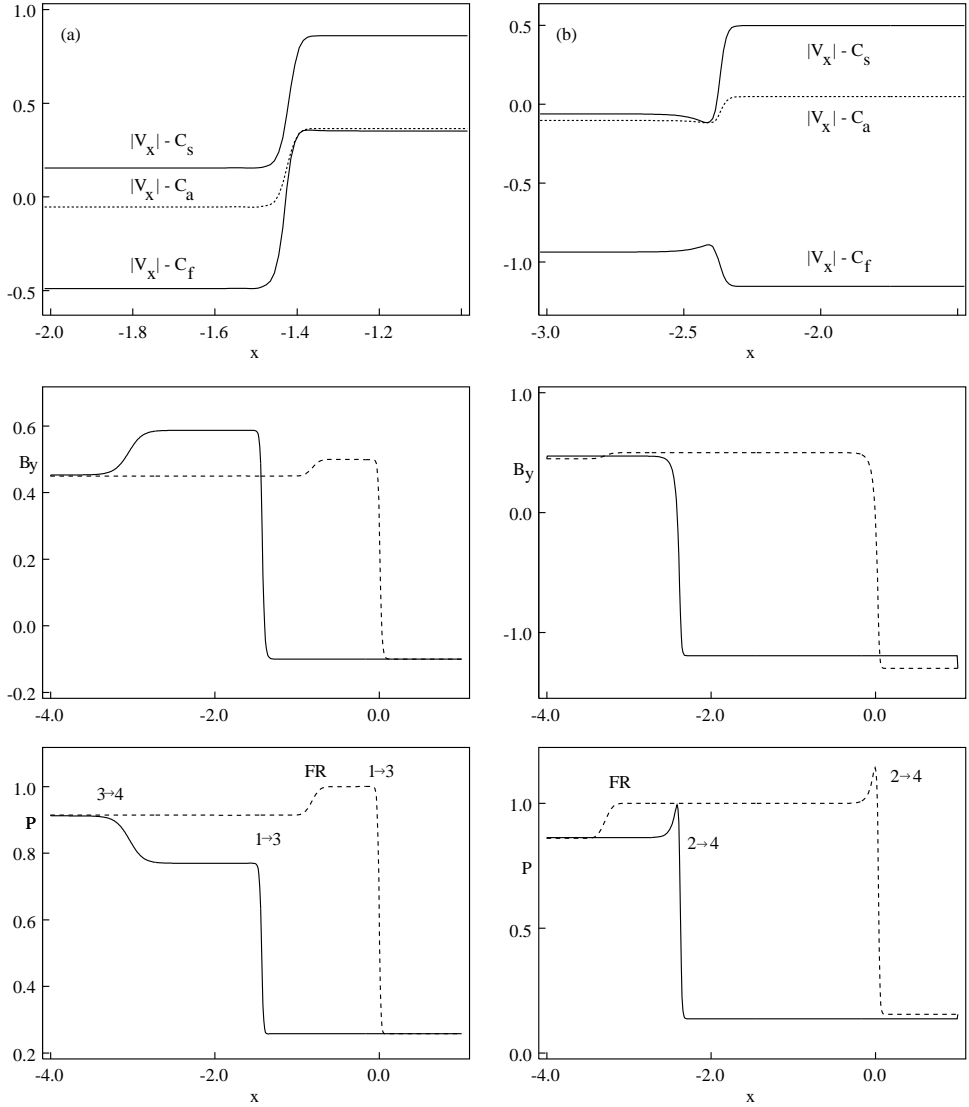}}
\caption{Planar simulations of the interaction between evolutionary shocks
and small amplitude fast rarefactions ($\delta B_t = 10\%$).  (a) fast ($1
\rightarrow 3$) shock, (b) slow ($2  \rightarrow 4$) shock.  In both cases
the outcome is  a shock of the same  type, together with some other waves.
Here  FR   denotes a fast  rarefaction  and  $V_x$  is the  x-component of
velocity as  measured  in the shock frame.     The dashed lines  show  the
initial   solutions.   The continuous  and   dotted  lines show the  final
solutions.}
\end{center}
\label{evsh}
\end{figure}
%fffffffffffffffffffffffffffffffffffffffffffffffffffffffffffffff

% --------------------------
\subsection{Planar MHD}
% --------------------------

We start  by discussing the results  of the planar simulations.  They show
that if the initial discontinuity corresponds to a  slow planar shock then
a smooth steady shock  structure  connecting the  initial left  and  right
states finally develops and  it does not  matter  whether the shock is  $3
\rightarrow 4$  or $2\rightarrow 4$.  The same  thing happens for the fast
planar shock and   the overdetermined (overcompressive) $1 \rightarrow  4$
shock.  In  contrast,  figure 2 shows  that  $2 \rightarrow 3$  shocks and
alfv\'en  shocks always turn   into  a slow  compound wave.   All  this is
exactly  as  perdicted by   the theory  described in   \S\ref{dis-str} and
\S\ref{apple}.  Our simulations cannot be used  to determine whether limit
shocks,  such $1  \rightarrow s$ and   $f \rightarrow  3$),  have a steady
dissipative shock structure,  simply because it is  impossible to set up a
shock whose speed is exactly equal to a characteristic speed.  However, if
we compute a Riemann problem that corresponds to a compound wave of any of
the types discussed above, the wave that is expected, or strictly speaking
a solution  close  to  such  a   wave, always  emerges.   This  is  hardly
surprising because  all  of  them have neighbouring   solutions containing
shocks with a steady dissipative structure.
        
%fffffffffffffffffffffffffffffffffffffffffffffffffffffffffffffff
\begin{figure} 
\begin{center} 
\leavevmode 
\hbox{% 
\epsffile[15 13 380 432]{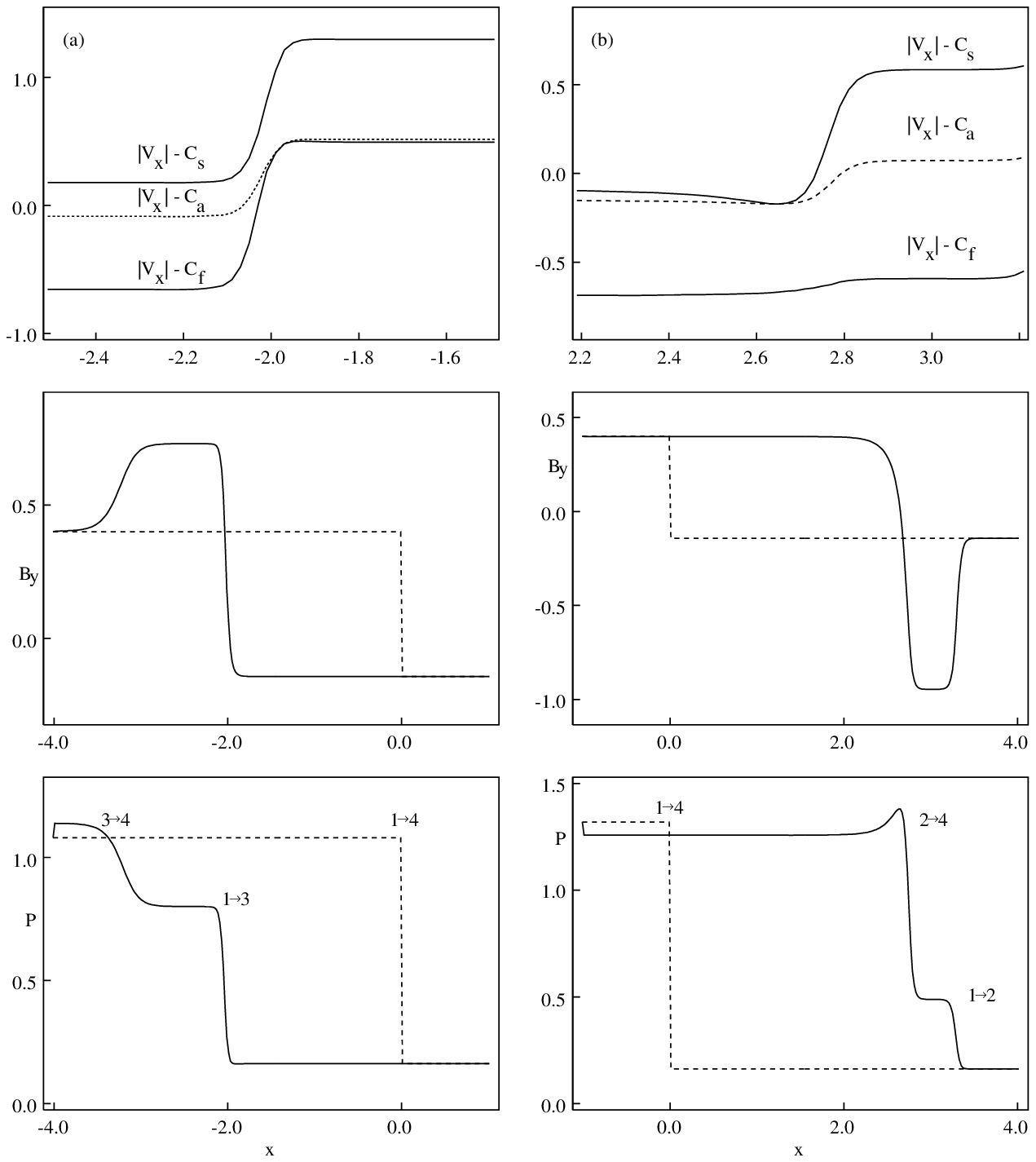}}

\caption{Planar  simulations of $1 \rightarrow   4$  shock subjected to  a
small variation of pressure ($\pm 10\%$) in  the left state. This shock is
non-evolutionary  even in planar  MHD  and splits  as  the  result of  the
perturbation into two evolutionary shocks plus other small amplitude waves
The outcome  is (a) $1  \rightarrow 3$  and  $ 3 \rightarrow  4$ shocks if
$\delta p = -10\%$ and (b) $1 \rightarrow 2$ and  $2 \rightarrow 4$ shocks
if $\delta p = +10\%$.  The dashed lines show  the initial solutions.  The
continuous and  dotted   lines show the   final  solutions.  $V_x$  is the
x-component of   velocity   as measured  in   the  frame of  the   emerged
intemediate shock.}
\end{center}
\label{is14}
\end{figure}
%fffffffffffffffffffffffffffffffffffffffffffffffffffffffffffffff
 
As is shown in  figure~3, evolutionary shocks always  survive interactions
with small  amplitude  waves  and  persist  if   the  Riemann  problem  is
perturbed.  Figure~4 shows  how  a small  variation  of  the initial  data
forces   an overdetermined  $1 \rightarrow 4$   shock  to  split into  two
evolutionary shocks.  Depending on the form of the perturbation, the shock
either splits into a $1 \rightarrow 2$ shock followed  by a $2 \rightarrow
4$ shock or a  $1  \rightarrow 3$ shock followed  by  a $3 \rightarrow  4$
shock.  This is to be expected because, as one can see from figure~1, a $1
\rightarrow 4$ shock is exactly equivalent to one  or other of these shock
pairs propagating with   the same  speed.   In  fact,  this  result is  in
complete agreement with the analysis of the Riemann problem for planar MHD
in  Myong and Roe  (1997b).  $1  \rightarrow 4$  shocks, {\it O-shocks} in
their notation, are only required on the boundary  between the two domains
of parameter space in which their solution  involves a combination of fast
and slow planar shocks (S2 and S1).

The results for    compound waves involving   non-evolutionary  shocks are
similar.  Figure~1 shows that the non-evolutionary $1 \rightarrow s$ limit
shock   can be understood as  a  {\it double-layer} shock  composed of two
evolutionary shocks, a $1 \rightarrow 2$ and a $2 \rightarrow s$.  Indeed,
if the Riemann problem corresponding to a  compound wave containing such a
shock is perturbed, then in some cases the outcome  is a $1 \rightarrow 2$
shock and a slow compound wave and  in other cases  it is a $1 \rightarrow
3$ shock and a detached slow rarefaction.

{\it All this can  be summed up by saying  that in planar MHD the behavior
of shocks  in our  numerical simulations is   entirely consistent with the
classical   evolutionary theory of  shocks  and  the theory of dissipative
shock structures as described in \S\ref{sgeneral} and \S\ref{dis-str}.}

% ------------------------
\subsection{Full MHD}
% ------------------------

Since both fast  ($1 \rightarrow 2$) and  slow ($3 \rightarrow  4$) shocks
satisfy the strong evolutionary condition in full MHD they are expected to
have  unique  dissipative structure  and be stable   with respect to small
perturbations of  any  kind.  This   is precisely what   we  find from our
simulations.

$1 \rightarrow 3$ and $2 \rightarrow  4$ shocks are overdetermined in full
MHD and it is therefore  possible that they might  have a nonunique steady
dissipative structure, indeed it turns out that they do.  These shocks, as
well as $1 \rightarrow 4$ shocks,  can now have a nonvanishing z-component
of magnetic  field inside the  shock layer even  if $B_z=0$  outside.  For
given the dissipative coefficients  their stucture can be parameterised by
the value of the following integral
\[
   I_z = \int_{-\infty}^{+\infty} B_z dx . 
\]
We can gradually increase  or decrease the value of  $I_z$ by sending from
the downstream side of the shock  an alfv\'en wave  that first rotates the
magnetic  field  by a small angle  and  then restores  the original state.
This wave is absorbed by the  shock which develops  a new steady structure
(see Figure~5{\it  a}).  However, like Kennel  et al. (1990) we found that
there is  a maximum value of  $|I_z|$ that the shock  can manage.  If this
limit  is exceeded,  then  the shock disintegrates  (see  Figure~{\it b}).
This does not  occur in  the case of   fast and  slow shocks because   the
alfv\'en  waves do not  get trapped  inside the  shocks, but  instead pass
straight through.

$2 \rightarrow 3$ shocks have the right number of incoming characteristics
and may therefore have a unique dissipative  structure in full MHD.  Since
such a structure does not exist in planar MHD,  we can only expect to find
them in our  simulations  by allowing a non-zero  $B_z$.   In order  to do
this, we  modified the initial  data by  inserting  a layer in  which  the
transverse field rotates smoothly from that in  the original left state to
that in  the original  right  state.   We found  that  the solution  never
relaxed to a smooth steady $2 \rightarrow  3$ transition and were about to
conclude   that no steady structure   exists  until we  realised that  the
solution shown  in figure~5{\it b}  actually contains a  $2 \rightarrow 3$
shock, which was  produced by the disintegration of  the $1 \rightarrow 3$
shock.  We therefore we studied the reaction  of a $1 \rightarrow 3$ shock
to an increase in $I_z$.   After absorbng another  alfv\'en wave the shock
splits and one  of the emerging  waves is again a  $2 \rightarrow 3$ shock
but of smaller amplitude  (figure~6).   This behaviour is consistent  with
the existence  of a  unique dissipative structure   for $2  \rightarrow 3$
shocks.   In fact, what happens  is  that, as  $I_z$ increases, the  shock
tends to an alfv\'en shock that rotates the transverse field by $\pi$

Finally, we have also verified  that all intermediate shocks and  compound
waves disintegrate when exposed to perturbations that  render the left and
right states non-coplanar.  For example, figure~7 shows how $1 \rightarrow
3$ and $2   \rightarrow  4$ shocks  split into  evolutionary  waves  after
interaction with a small amplitude alfv\'en wave.  After the alfv\'en wave
has been absorbed the transverse fields on either side of the shock are no
longer parallel  or antiparallel as required  by the shock equations.  The
shock can  only become coplanar by  emitting alfv\'en waves, which, for an
intermediate shock,    can only  be  done  in   the downstream  direction.
However, since there  is no downstream  travelling alfv\'en  wave that can
restore  the  original  post-shock state,  the shock    must  split.  This
argument is not new, in fact it was used  by Kantrowitz \& Petschek (1966)
to prove that intermediate shocks are  unphysical.  The wave designated as
AW in figure~7 can be called a dissipative alfv\'en wave but it could also
be  described  as an  evolving $2 \rightarrow   3$ shock  with a gradually
increasing value of $I_z$.

{\it We therefore conclude that for full MHD the behaviour of shocks in our
numerical simulations  is also   entirely consistent with    the classical
evolutionary   theory of  shocks   and  the  theory  of dissipative  shock
structures as described in \S\ref{sgeneral} and \S\ref{dis-str}.}

%ffffffffffffffffffffffffffffffffffffffffffffffffffffffffffffffffffffff
\begin{figure} 
\begin{center} 
\leavevmode 
\hbox{% 
\epsffile[15 13 380 289]{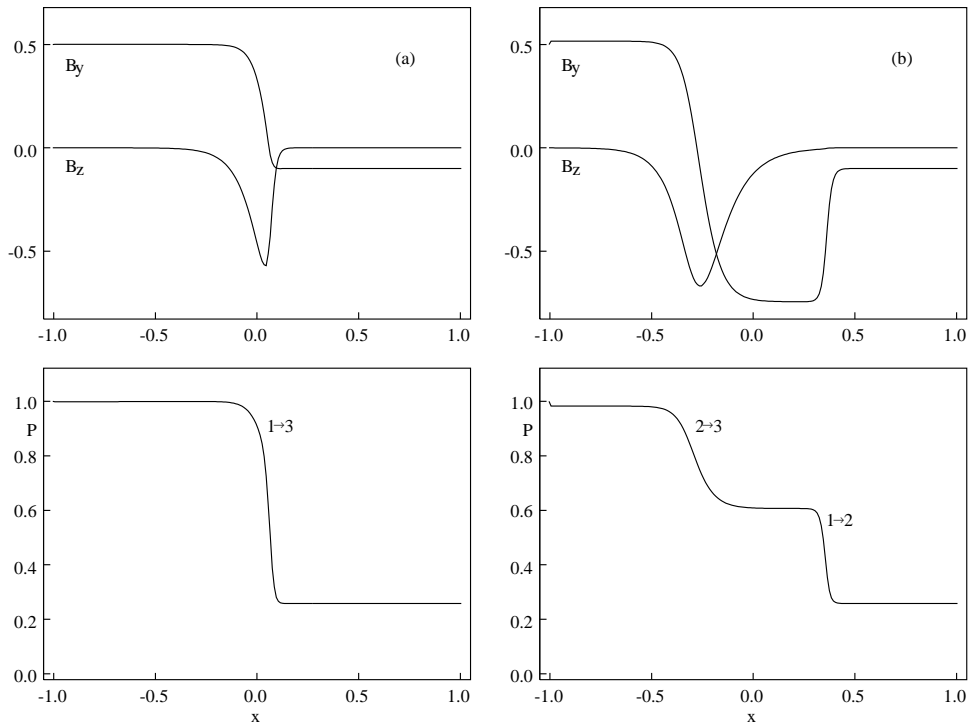}}

\caption{Dissipative structure of a $1  \rightarrow 3$  shock in full  MHD
for  different  values of   $I_z$.  This   shock has  a  nonunique  steady
dissipative structure that depends upon $I_z$. For relatively small values
of $I_z$  this structure  is steady  (left  panel, $I_z=-0.085$)  but  for
larger values it   splits into $1  \rightarrow  2$ and $1   \rightarrow 2$
shocks (right panel, $I_z=-0.20$.}

\end{center}
\label{is13}
\end{figure}
%ffffffffffffffffffffffffffffffffffffffffffffffffffffffffffffffffffffff

%ffffffffffffffffffffffffffffffffffffffffffffffffffffffffffffffffffffff
\begin{figure} 
\begin{center} 
\leavevmode 
\hbox{% 
\epsffile[15 13 380 195]{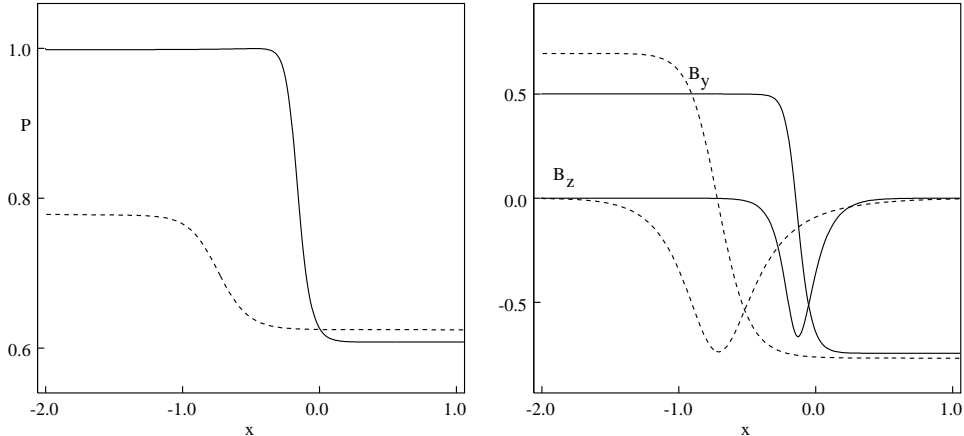}}
\caption{Dissipative structure of  a $2 \rightarrow  3$ shock in full MHD.
This shock has a unique steady dissipative  structure and therefore reacts
to a change in $I_z$ by  emitting some waves and  turning into a different
$2 \rightarrow  3$ shock.   The  continuous lines  show the   solution for
$I_z=-0.20$ and the dashed lines for $I_z=-0.52$.}
\end{center}
\label{is23}
\end{figure}
%ffffffffffffffffffffffffffffffffffffffffffffffffffffffffffffffffffffff

%ffffffffffffffffffffffffffffffffffffffffffffffffffffffffffffffffffffff
\begin{figure} 
\begin{center} 
\leavevmode 
\hbox{% 
\epsffile[15 13 380 445]{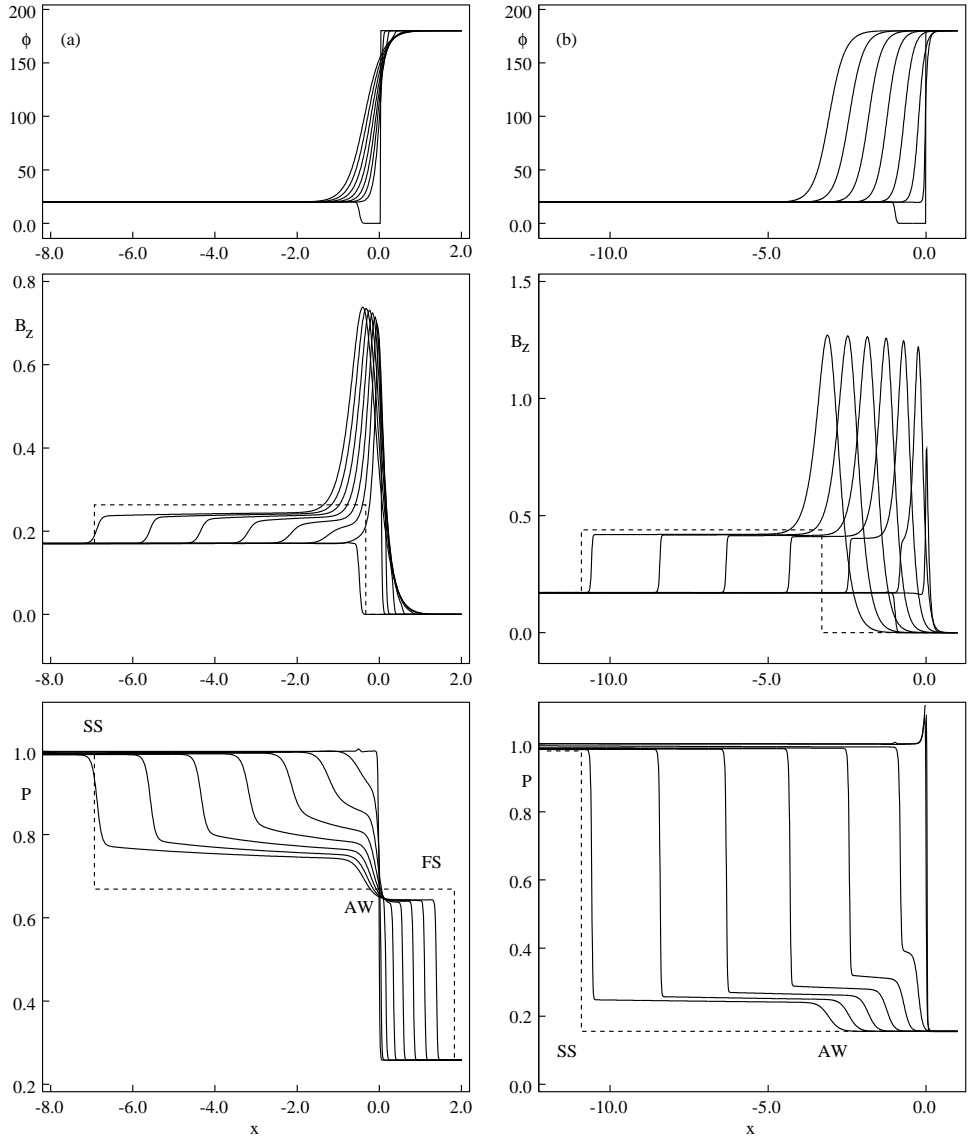}}
\caption{MHD shocks interacting with  small amplitude alfv\'en waves.  The
evolutionary  shocks survive, but  the non-evolutionary ones split.  (a) A
$1 \rightarrow  3$ shock splits into a  fast shock  (FS), an alfv\'en wave
(AW)  and a slow shock (SS);  b) A $2  \rightarrow 4$ shock splits into an
alfv\'en   wave and a   slow shock. Other  small  amplitude waves are also
emitted.  The dashed  line shows the  exact ideal solution of  the Riemann
problem for the initial state formed  by the collision of the intermediate
and alfv\'en shocks.}
\end{center}
\label{sinta}
\end{figure}
%ffffffffffffffffffffffffffffffffffffffffffffffffffffffffffffffffffffff

%ffffffffffffffffffffffffffffffffffffffffffffffffffffffffffffffffffffff
\begin{figure} 
\begin{center} 
\leavevmode 
\hbox{% 
\epsffile[0 0 390 357]{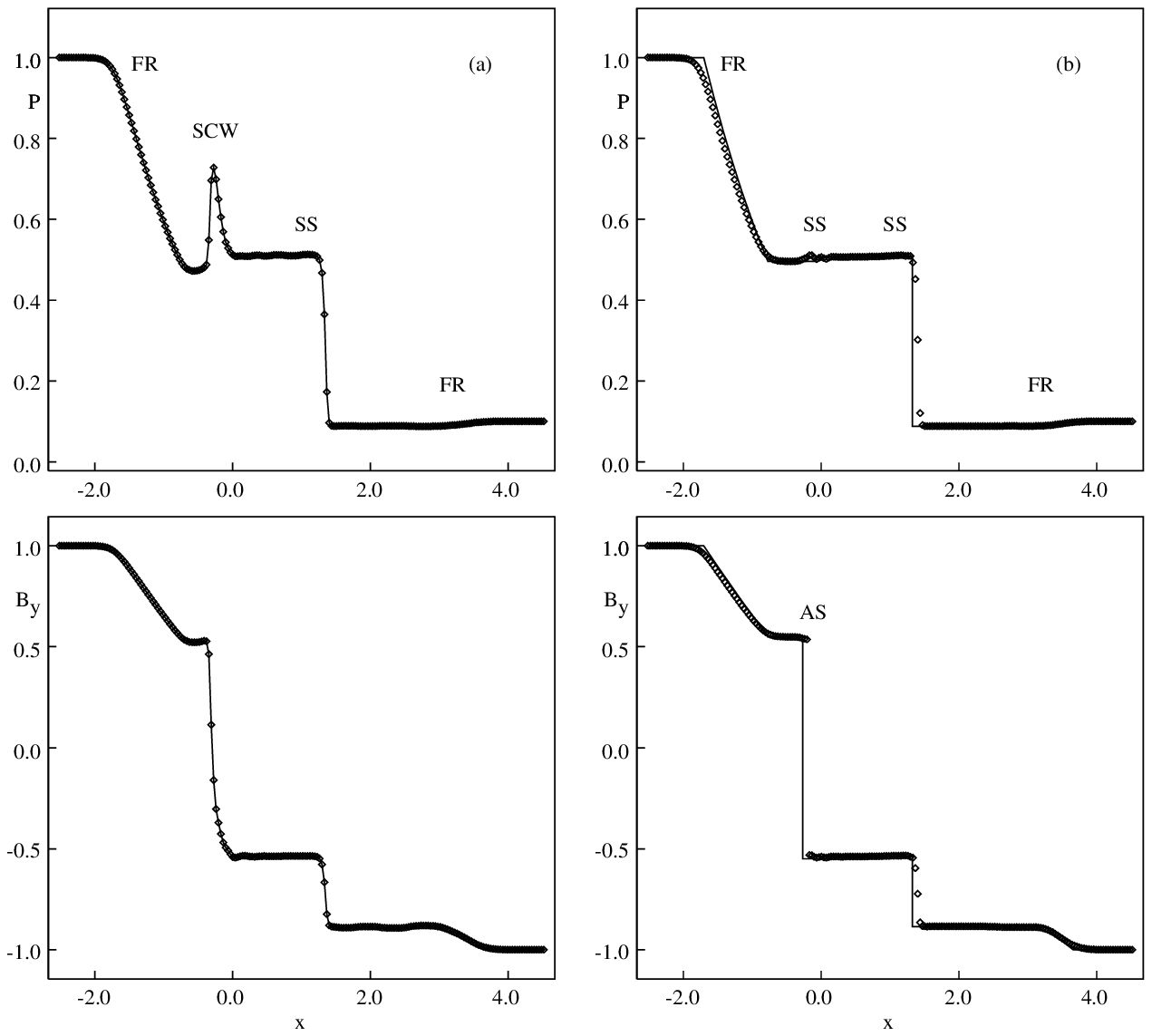}}
\caption{Brio  \& Wu problem  (Brio \&  Wu  (1988). (a) Numerical solution
found  using a  Godunov  type scheme. This  is  a  proper solution  of the
reduced system  of  planar  MHD but is  inadmissible  in  full  MHD.   (b)
Numerical solution   found    using Glimm's  scheme  to  track    alfv\'en
discontinuties    (markers) and    the   exact solution    involving  only
evolutionary shocks (lines). This is a proper solution for full MHD and is
the only physically admissible solution for this problem.}
\end{center}
\label{brio}
\end{figure}
%ffffffffffffffffffffffffffffffffffffffffffffffffffffffffffffffffffffff

% ===============================
\section{Discussion} 
\label{dissc}
% ===============================

The results described in the previous sections have clarified many aspects
of the shock theory in general and MHD shocks  in particular and provide a
basis upon which we can discuss other important, related, issues.

\subsection{Riemann problems and evolutionary conditions} 

One of the arguments in favour of  non-evolutionary shocks used in current
literature is  that some Riemann  problems do not   have a solution unless
non-evolutionary shocks are  admitted  (e.g.  Glimm   1988, Myong  \&  Roe
1997a,b).  This is presumably based on the belief that any Riemann problem
must have a  physically admissible solution.   Although this is  certainly
true for gas dynamics, there is surely no reason why this has to hold for any
system.  It all comes  down to the notions  of bifurcations and structural
stability.  One has to  ask the following question:  is it, or is  it not,
possible to  carry out the  relevant experiment in  a laboratory?   If the
qualitative result   of the experiment  does not  change when  the initial
conditions are slightly  changed, then the  problem is structurally stable
and the experiment is possible, at least in  principle.  However, if this
is not true, then the problem is  structurally unstable and no appropriate
experiment is possible.  It therefore follows that the set of structurally
unstable Riemann problem are confined to  regions of parameter space whose
total  volume is zero.   Now suppose there  is an MHD Riemann problem that
has  no other solutions   than  those containing non-evolutionary  shocks.
Since there are arbitrary small perturbations of the parameters that cause
these shocks to split into evolutionary  shocks, this Riemann problem must
be  structurally unstable.  In  full MHD the only known   case for which a
non-evolutionary shock, a $1 \rightarrow 4$ shock, is required is a piston
problem in which  the piston velocity is   parallel to the magnetic  field
(Jeffrey \&  Taniuti  1964).  If this  condition  is not exactly satisfied
then the  non-evolutionary shock does not  arise.  Close inspection  of the
solution of the Riemann  problem for planar MHD presented  by Myong \& Roe
(1997b)  shows  that  non-evolutionary  shocks are   required only on  the
boundaries   between  domains  in   parameter  space  that   contain  only
evolutionary shocks.

\subsection{Steepening of continuous waves} 

Another  argument that appears  to  justify the existence of  intermediate
shocks is  based on the  results  of numerical simulations  by Wu  (1987),
which  suggest that intermediate    shocks   can be formed  by   nonlinear
steepening  of simple magnetosonic  waves.  Since the transverse component
of magnetic  field changes sign across  an  intermediate shock  the simple
wave must   have  the same    property, which  means  that the  transverse
component  of    magnetic field must   vanish somewhere   within the wave.
However, at this  point  the magnetosonic speed  is equal  to the alfv\'en
speed and  it is impossible to assign  a unique eigenvector to  the simple
wave.  As  the result, the  direction  of the  tangential component of the
field can rotate by an arbitrary  angle at this  point so that simple wave
really consists of two  distinct parts, which  are disconnected as far  as
the direction of  the magnetic field  is concerned.  This  can be put in a
slightly different way.  Alfv\'en waves propagating  in the same direction
as such a simple wave cannot pass throught the alfv\'en point.  During the
steepening they will accumulate near this point giving rise to a net field
rotation so that the discontinuity  that  forms has non-coplanar left  and
right states and  can therefore not be a  single shock.   Instead, it must
split  into  evolutionary   shocks, one  of  which  must   be an  alfv\'en
shock. Incidentally, this seems to be  the only way of generating alfv\'en
shocks.

However, in  planar  MHD the transition  through  the alfv\'enic  point is
unique and as we have  seen some of  the intermediate  shocks are in  fact
evolutionary.   This is  the explanation  for   the outcome of  the planar
simulations performed by Wu  (1987).  He also  found that the results were
not very  different if the initial data  was perturbed so  that  it was no
longer   exactly  coplanar.  However, because    of  the periodic boundary
conditions used  in this simulation,   there was no   net rotation in  the
perturbed problem, which makes it rather  artificial.  The reason why this
perturbation did not destroy the intermediate shock is that these boundary
conditions, together with the initial data, only  allowed a small value of
$I_z$  per shock.  It is therefore  hardly surprising that an intermediate
shock appeared since, as we have shown, these shocks  can survive if $I_z$
is small enough.
 
\subsection{Timescale for disintegration} 

Let us suppose that an intermediate shock has somehow been formed and then
interacts with an alfv\'en wave that rotates the magnetic field by a small
angle $\delta\phi$. It is clearly  of some importance  to know how long it
takes for  the shock to split.  Our  simulations show that it  splits when
the   the  value of $I_z$   associated with  the  shock  structure becomes
comparable with $lB_y$, where $l$ is the  shock thickness. If the incident
alfv\'en wave has a small amplitude, $\delta \phi$, then this gives us the
following estimate for the disintegration time, $t_s$

\begin{equation}
t_s \approx {l \over {c_a \delta\phi}},
\label{46}
\end{equation}

\noindent
where  we have used the alfv\'en  speed as a characteristic fluid velocity
in the shock frame.  This also tells us that the shock will only propagate
for a distance $\approx l/\delta\phi$ before it falls apart.  From this we
conclude that in all cases for which the dissipative scale is much smaller
then the characteristic length scale  of the flow, intermediate shocks can
only appear as very short lived time-dependent phenomena.

It  is  instructive to  apply   equation (\ref{46}) to  the interplanetary
intermediate shock for   which Chao et  al.   (1993) claim  to  have found
evidence in the Voyager 1 data.  In this case $c_a  = 40 {\rm ~km~s^{-1}}$
and   $l  = 5  \times  10^4  {\rm  ~  km}$,  which gives  $t_s  = 1.2 10^3
{\delta\phi}^{-1}{\rm~s}$.  The flow  time   for the  solar  wind at  this
distance ($\simeq 9 {\rm ~AU}$) is $\simeq 3  \times 10^7 {\rm ~s}$. It is
therefore clear that $\delta \phi$ would have to be ridiculously small for
the shock to survive  for a significant fraction of  a flow time.  This is
most unlikely since the flow of the solar  wind is sufficiently complex to
contain plenty of alfv\'en waves for which $\delta \phi \sim 1$ and indeed
Chao et al.  find   plenty of evidence for  strong  alfv\'en waves  in the
data.  Actually, the evidence for an intermediate shock is not really very
convincing.   The uncertainties are such  that it could just  as well be a
slow shock.

Exactly  the  same  arguments can  be   applied to the magnetohydrodynamic
shocks in   the interstellar   medium.   Not  only  does  the  theory   of
collisionless shocks (see e.g.  Tidman   \& Krall 1971) predict that,   in
these conditions, such shocks are extremely  thin compared to the scale of
the  flow but there  are numerous  observations  that confirm that this is
indeed true (see e.g. Draine \& McKee 1993).

\subsection{Convexity of MHD}  

From  the above discussion  it is quite clear  that a hyperbolic system is
genuinely non-convex if it allows structurally  stable compound waves that
only contain  evolutionary shocks.  Planar    MHD is therefore   genuinely
non-convex whereas full MHD is convex.

\subsection{Non-evolutionary shocks in numerical simulations}  

The appearance of non-evolutionary shocks in numerical calculations is not
something that is unique to  MHD since it is well  known that, even in gas
dynamics, some numerical schemes  can generate expansion shocks in certain
circumstances.   However, this  phenomenon is  both  more subtle  and more
interesting in the  case of MHD.  The essential  point is that, unlike gas
dynamics, planar MHD is is very different from full  MHD in the sense that
there are shocks that  are non-evolutionary in  full MHD, but evolutionary
in planar MHD and vice-versa.  Unfortunately, this property means that the
results of planar MHD simulations can be very misleading because, although
most upwind schemes seem to give perfectly good  solutions for planar MHD,
these  are of no  relevance to  the real universe  with its  three spatial
dimensions.   This is not at  all unusual, indeed  it may very well be the
rule rather  than the exception.   For example,  the properties  of  fluid
turbulence are very different in two and  three dimensions as are those of
magnetohydrodynamic dynamos.

The other properties  of non-evolutionary MHD shocks,  that are not shared
by gas dynamical expansion shocks, are that all of them satisfy the second
law of thermodynamics  and most of them also  possess a steady dissipative
structure.  This, together  with the fact that  the ratio of the thickness
of numerical shock structures  to the overall scale of  the flow is almost
always many orders of magnitude greater than in the corresponding physical
system, means that   they  can persist   for a significant   time  even in
nonplanar  problems.   For example,  if  the piston  problem  discussed by
Jeffrey \& Taniuti (1964, p. 256--258) is slightly modified so that it has
a small transverse component of  the field, then the evolutionary solution
contains fast, slow and alfv\'en shocks  all propagating with very similar
speeds. In a numerical simulation this complex would remain unresolved for
some  time, during  which it would  be  classified as a  $1 \rightarrow 4$
shock.

The only  truly satisfactory  solution to   this difficulty is  to  devise
schemes that only allow evolutionary shocks. Figure~8 shows that there are
schemes that will do this.  Here we have a  numerical solution to the Brio
\& Wu problem  obtained with  our  MHD version  of Glimm's scheme   (Glimm
1965).  This method requires a nonlinear Riemann  solver and we employ the
one   described   in Falle et   al.   (1998),  which specifically excludes
intermediate shocks.  In fact we do not use Glimm's scheme everywhere, but
only  to track the  alfv\'en shock.  One can see  that  in this way we can
avoid  the  appearance of  intermediate  shocks  even in  planar problems.
Unfortunately, it is  not a simple matter  to generalise this to more than
one dimension.

The only viable option, that we can think of, is  to subject all numerical
calculations  to  a careful  analysis using the   theory described in this
paper.  As an example  of  this, it  is instructive   look at  some recent
calculations of steady MHD flow past a cylinder.

\subsection{2D bow shock simulations} 

Recently,   De Sterck et    al.(1998)    have carried out numerical    MHD
calculations of the flow  past an infinite, perfectly conducting cylinder.
These are  planar simulations and  must therefore  be  interpreted in  the
light of the theory of planar MHD.  The parameters  are chosen in such way
that the usual convex bow shock is impossible. Instead, the analysis given
in Steinolfson \& Hundhausen (1990) suggests that the  shock has a dimple.
They assumed that there is only a single  shock, in which case a cosistent
solution requires the shock  type to change from $1  \rightarrow 2$ to  $1
\rightarrow  3$ and then to $1   \rightarrow 4$ as  the  distance from the
symmetry   axis  decreases.  Although  the   $1  \rightarrow 4$  shock  is
non-evolutionary  even in planar  MHD, in this case  it  seems that such a
shock must occur on the symmetry  axis for the same  reason that it occurs
when a  piston  moves parallel to the  magnetic  field. However, one would
expect it to split  into $1  \rightarrow 2$ and   $2 \rightarrow 4$ or  $1
\rightarrow  3$ and $3   \rightarrow 4$ shocks   further away from the the
axis.  Indeed, De Sterck et al.(1998) find that  not far from the axis
the $1 \rightarrow   4$ shock splits and  the  leading shock (ED in  their
notation) is a  $1 \rightarrow 2$.  At  some distance from this  branching
point the other shock (EG) is identified by them as $f \rightarrow s$, but
this  is unlikely to  be true everywhere for  such  an inhomogeneous flow.
One would also expect another branching  at the point where Steinolfson \&
Hundhausen (1990)  predict  a  transition from   $1 \rightarrow  3$  to $1
\rightarrow  4$.  The results of De  Sterck  et al.(1998) do, indeed, show
this  branching   (DE and  DG),   with the   trailing shock  being clearly
identifiable as a $2 \rightarrow 4$ shock.

%xxxxxxxxxxxxxxxxxxx

%================================================================
\section{Conclusions}
\label{conc}
%================================================================

Both   our analysis and   numerical   results show that the   evolutionary
conditions for existence and  uniqueness of discontinuous solutions of the
equations of ideal MHD  are not  only compatible  with the  conditions for
existence and uniqueness of steady  dissipative shock structures, they are
actually complementary to them.   The  general theory suggests that   this
will be true  for all  nonlinear hyperbolic  systems   that can arise   in
nature.  Non-evolutionary shock can have a nonunique dissipative structure
and may, perhaps, appear under  some exeptional curcumstances as transient
phenomena.  However,  they are not persistent and  are bound to split when
subjected to small perturbations.  In the case  of MHD, alfv\'en waves are
the most effective killers since not only  our calculations but also those
described by Wu (1988a) show that intermediate MHD shocks are destroyed by
interactions with alfv\'en waves.  It is true that it  takes a finite time
for  this interaction to  take place, but  in  any physical system that we
know of, this time is so short that  it is most  unlikely that such shocks
can be detected.

The occurrence of intermediate MHD  shocks in planar numerical simulations
is consistent with the mathematical properties  of planar MHD, in which $1
\rightarrow 3$  and $2 \rightarrow 4$ shocks   become evolutionary but the
alfv\'en shock becomes non-evolutionary.  However,  the planar limit is  a
singular limit  of  full   MHD and we     suggest that  planar   numerical
simulations should be  avoided,   especially  since they are    hardly any
cheaper than for full MHD.
 
Intermediate shocks   may  even pollute   full MHD   simulations   because
numerical  shock structures  are  usually not very  thin  compared  to the
length scale of  the flow.  It is therefore  essential that the results of
such simulations be subjected to a careful analysis  in order to make sure
that they do   not contain any intermediate   shocks.   If they  do,  then
additional  work is required  to  determine the extent   to which they are
corrupted.  The results of our  calculations with Glimm's scheme show that
this problem can   be eliminated in   numerical schemes that  treat shocks
especially alfv\'en shocks, as discontinuities.

\end{document}